%% file: main.tex
\documentclass[sigconf]{acmart} % for ACSAC
\usepackage{soul}
\usepackage{makecell}

\AtBeginDocument{%
  \providecommand\BibTeX{{%
    \normalfont B\kern-0.5em{\scshape i\kern-0.25em b}\kern-0.8em\TeX}}}

\setcopyright{acmcopyright}
\copyrightyear{2021}
\acmYear{2021}
\acmDOI{10.1145/1122445.1122456}

\acmConference[Redacted Conference]{Redacted Conference 2021}{December, 2021}{Somewhere on Earth}
\acmBooktitle{Proceedings of the Redacted Conference}
\acmPrice{15.00}
\acmISBN{978-1-4503-70899}

\begin{document}

\title{Characterizing Malicious URL Campaigns}

\author{Mahathir Almashor}
\email{mahathir.almashor@data61.csiro.au}
\authornote{Commonwealth Science and Industrial Research Organisation (CSIRO) Data61}
\authornote{Cyber Security Cooperative Research Centre}
\affiliation{}
% \affiliation{\institution{CSIRO's Data61 \& Cyber Security Cooperative Research Centre}}

\author{Ejaz Ahmed}
\email{ejaz.ahmed@data61.csiro.au}
\authornotemark[1]
\affiliation{}
% \affiliation{\institution{CSIRO's Data61}}

\author{Benjamin Pick}
\email{benjamin.pick@data61.csiro.au}
\authornotemark[1]\authornotemark[2]
\affiliation{}

\author{Sharif Abuadbba}
\email{sharif.abuadbba@data61.csiro.au}
\authornotemark[1]\authornotemark[2]
\affiliation{}

\author{Raj Gaire}
\email{raj.gaire@data61.csiro.au}
\authornotemark[1]\authornotemark[2]
\affiliation{}

\author{Shuo Wang}
\email{shuo.wang@data61.csiro.au}
\authornotemark[1]\authornotemark[2]
\affiliation{}

\author{Seyit Camtepe}
\email{seyit.camtepe@data61.csiro.au}
\authornotemark[1]\authornotemark[2]
\affiliation{}

\author{Surya Nepal}
\email{surya.nepal@data61.csiro.au}
\authornotemark[1]\authornotemark[2]
\affiliation{}

% \affiliation{%

\renewcommand{\shortauthors}{Almashor et al.}
\newcommand{\sharif}[1]{\textcolor{blue}{[Sharif: #1]}}
\newcommand{\ejaz}[1]{\textcolor{red}{[Ejaz: #1]}}

\input{Sections/0_Abstract}
\maketitle

\input{Sections/1_Introduction}

\input{Sections/2_Background}

\input{Sections/3_Individual_submission_char}

\input{Sections/4_Campaign_Identification}

\input{Sections/5_Case_studies}

\input{Sections/7_Discussion}

\input{Sections/8_Related_work}
\input{Sections/9_Conclusion}

%\bibliographystyle{ACM-Reference-Format}
\bibliographystyle{unsrt}
\bibliography{references}

\end{document}

%% file: Sections/0_Abstract.tex
\begin{abstract}
URLs are central to a myriad of cyber-security threats, from phishing to the distribution of malware. Their inherent ease of use and familiarity is continuously abused by attackers to evade defences and deceive end-users. Seemingly dissimilar URLs are being used in an organized way to perform phishing attacks and distribute malware. We refer to such behaviours as \textit{campaigns}, with the hypothesis being that attacks are often coordinated to maximize success rates and develop evasion tactics. The aim is to gain better insights into campaigns, bolster our grasp of their characteristics, and thus aid the community devise more robust solutions. To this end, we performed extensive research and analysis into 311M records containing 77M unique real-world URLs that were submitted to VirusTotal from Dec 2019 to Jan 2020. From this dataset, 2.6M suspicious campaigns were identified based on their attached metadata, of which 77,810 were doubly verified as \textit{malicious}. Using the 38.1M records and 9.9M URLs within these malicious campaigns, we provide varied insights such as their targeted victim brands as well as URL sizes and heterogeneity. Some surprising findings were observed, such as detection rates falling to just 13.27\% for campaigns that employ more than 100 unique URLs. The paper concludes with several case-studies that illustrate the common malicious techniques employed by attackers to imperil users and circumvent defences.
\end{abstract}

%% file: Sections/1_Introduction.tex
\section{Introduction}\label{sec:intro} 
Regardless of their intent, malicious actors have relied on the humble Uniform Resource Locator (URL) as the penultimate step in their pernicious operations. Littered throughout phishing emails, social network spam, and suspicious web-sites, these otherwise common text strings are crafted to mislead end-users \cite{althobaiti2021idontneed, chiba2019domain}. This results in the divulging of sensitive personal and/or commercial information via fake login pages~\cite{maroofi2020areyou}, or the inadvertent download of malware which compromise machines and allow unauthorized access \cite{shibahara2017detecting}.

A wide body of work exists on the urgent issue of harmful URLs, from detection at the domain name level \cite{zhauniarovich2018asurvey, hu2021assessing, desmet2021premadona} to deep-learning classification of entire URL strings \cite{das2020deep, sameen2020phishhaven}. Yet, rarely have URLs been used in an isolated fashion. Attackers incorporate them into multi-stage compromise strategies, with reliance on waves of unique URLs to maximise their eventual reach and evade cyber defences. Witness the recent surge in attacks based on the COVID-19 pandemic \cite{cyber2021covid19} that used carefully crafted URLs and phishing emails to deceive users. Criminal organisations often include popular brand and domain names in their malicious URLs to better target their intended victims (e.g., http://mail.get-apple[.]support). Such groups of URLs are what we define as \textit{malicious campaigns}.

The work herein shares the approach of prior campaign characterization efforts, including phone-based robocalls \cite{prasad2020whoscalling}, social network spam \cite{gao2010detecting} and domain names \cite{weber2018unsupervised}. In \cite{starov2018betrayed}, groups of websites (including malicious ones) are clustered via advertising tokens matched to the same owner. It is reported in \cite{oest2020sunrise} that a small percentage of operators were behind the most successful phishing campaigns, and that threats can persist for up to 9 months by employing various evasion tactics. As for the URLs themselves, prior focus has been on their general treatment as individual pieces. We instead seek to identify and characterize clusters of seemingly unique and disassociated URLs as a motivated whole and thus, gain more pertinent insights into such concerted attacks.

% \begin{table*}[t]
% % \footnotesize
% \centering

%     \begin{tabular}{c c c c l l}
%     \textbf{Scheme} & \textbf{Sub-domain} & \textbf{Domain} & \textbf{Suffix} & \textbf{Path \& Queries} & \textbf{Remarks}\\
%     \hline
%     https:// & help & abc & net.au & /hc/en-us/requests/new?ticket\_form\_id=360000036795 & benign\\
%     https:// & support & apple & com & /en-us/HT204306 & benign\\
%     http:// & mail & get-apple & support & /index.php & malicious\\
%     https:// & www.apple.com & online-support & in & /index.php/step2.php & malicious\\
%     %  \hline
%     \end{tabular}

% \caption{Components of a URL and examples seen in the collected dataset}
% \label{tab:urls}
% \end{table*}

This is exemplified in one of our key findings: Google Safe Browsing\footnote{https://safebrowsing.google.com/}, the default block-list used in most web-browsers, is not able to detect \textit{all} unique URLs for 49.76\% of 18,360 multi-URL malicious campaigns we discovered. Considering the expectation that malware is often unleashed in waves, this indicates that a more refined approach is warranted. The community needs insights into campaigns of malicious URLs, so as to analyze and learn the evasion and deception techniques that attackers exploit. This also throws open the possibility of timely detection of large-scale campaigns and informs potential mitigation strategies that can be employed.

We begin by pre-processing and analysing 311,588,127 submissions to the VirusTotal platform\footnote{https://www.virustotal.com}. These represented individual URL submissions from 2019-12-05 to 2020-01-31, with a total uncompressed size of close to 2TB. To the best of our knowledge, no prior work has attempted the analysis of a comprehensive dataset of such size stretched across a prolonged 2 month period. The work by Starov \textit{et al.} in \cite{starov2018betrayed} examined only 145K URLs over a duration of 2 weeks, while the raw work in \cite{le2018urlnet} used 15M URLs but treated the data as individual pieces for classification. Attempting work on such large scale presented many challenges but proved vital, as robust findings on the nature of malicious campaigns require high confidence in the completeness of records over the duration of the analysis. That is, the identification of campaigns and subsequent results is weakened should gaps exist in the historical data.

Given this novel perspective and the access to a large dataset, our research questions and contributions are summarised as follows:

% From this novel perspective on URLs, we deliver 22 key findings in response to the following research questions:

\begin{itemize}

    \item[\textbf{RQ1}] \textbf{What patterns can we discern from submissions over a prolonged period?} Our contributions begin with Section 3, where we deliver 10 key findings on the nature of submissions. We detail how, when and by whom URLs are submitted, delve into their general temporal characteristics as well as high-level vendor performance.
    
    % \item \textbf{What is the nature of URL submissions over a prolonged 2 month period?} Here, we seek to understand the general characteristics of VirusTotal submissions. What percentage are flagged as malicious by vendors? Which country creates the most submissions? The collected data also corresponds to a typical holiday period, with potentially interesting patterns to be discerned. Do malicious campaigns ramp up during the holiday period or enter a lull as offices around the world are shut down?
    
    \item[\textbf{RQ2}] \textbf{Can we cluster URLs into campaigns for subsequent characterization?} Section 4 details how we grouped seemingly divergent URLs by simply using content hash metadata available within submissions. We include 9 findings on the identification, risk-factors, verification and the nature of URLs within verified malicious campaigns. In-depth metrics are also provided along dimensions such as campaign size, domain names used and vendor detection rates.
    
    \item[\textbf{RQ3}] \textbf{What attack methods can we see in campaigns?} We continue with a detailed presentation of several case-studies in Section 5, which include investigations into targeted victim brands, payload types and fileless malware injected directly into the URL itself. The aim is to encapsulate the tactics and methodology of attackers as seen within the dataset and provide insights to the community as to the different types of malicious URLs and strategies employed.

\end{itemize}

To conclude the paper, we review a campaign where a new malicious URL could have been easily flagged by matching the content hash of a known sibling. This is followed by a discussion on the impact of the various findings presented, with further recommendations that aim to dispel common misconceptions of campaigns.

%% file: Sections/2_Background.tex
\begin{table}[b]
% \footnotesize
\centering

    \begin{tabular}{c c c c l l}

    \textbf{Scheme} &
    \textbf{Sub-domain} &
    \textbf{Domain} &
    \textbf{Suffix} &
    \textbf{Path}\\
    % \textbf{Remarks} \\
    
    \hline
    
    % https:// & help & abc & net.au & /hc/en-us/requests/new?ticket\_form\_id=360000036795 & benign\\

    https:// &
    support &
    apple &
    com &
    /en-us/ \\
    % benign\\

    http:// &
    mail &
    get-apple &
    support &
    /index.php \\
    % malicious\\

    % https:// & www.apple.com & online-support & in & /index.php/step2.php & malicious\\
    
    \end{tabular}

\caption{Components of URLs}
\label{tab:urls}
\end{table}

\newcolumntype{L}[1]{>{\centering\arraybackslash}m{#1}}
% \newcolumntype{A}[1]{>{\arraybackslash}m{#1}}
\begin{table*}[t]
\centering

    \begin{tabular}{%
        p{0.08\linewidth}|%
        p{0.62\linewidth}|%
        p{0.2\linewidth}%
    }
    % \begin{tabular}{|p{0.1\linewidth}|l|p{0.6\linewidth}|}
    % \hline

    \textbf{Field} & \textbf{Description} & \textbf{Sub-Fields \& Examples}
    \\[.5em]
    %--------------------------------
    % \hline
    url
    &
    The URL that was submitted for analysis
    &
    http://example[.]com
    \\[1.0em]
    %--------------------------------
    % \hline
    \makecell[l]{content\\hash}
    &
    \makecell[l]{%
        A SHA-256 hash of the \textit{content} that the URL was pointing to. For example, if the\\
        URL pointed to a file, that file would be downloaded and a hash derived from it.
    }
    &
    f568...215e
    \\[1.0em]
    %--------------------------------
    % \hline
    positives
    &
    \makecell[l]{%
        A count of how many vendors have flagged the URL, with $0 <= n < N$ where $N$\\
        is the total number of vendors.
    }
    &
    0
    \\[1.0em]
    %--------------------------------
    % \hline
    first\_seen
    &
    \makecell[l]{%
        The first time this particular URL was submitted to VirusTotal. There is also a\\ \texttt{scan\_date} field which signifies when the URL was scanned.
    }
    &
    2015-11-06 00:49:02
    \\[1.0em]
    %--------------------------------
    % \hline
    submission
    &
    \makecell[l]{%
        Information specific to the user or application that submitted the URL, including\\
        country of origin, date/time and by what method.
    }
    &
    \makecell[l]{%
        id:f69a77a0, country:ES,\\
        interface:api\\
        % date:2019-12-22 10:00:42\\
    }
    \\[1.0em]
    % %--------------------------------
    % \hline
    % scan\_id
    % &
    % \makecell[l]{%
    %     1dc4...8464-1577008842
    % }
    % &
    % \makecell[l]{%
    %     An identifier for the submission comprising of a SHA-256 representation of the URL\\
    %     string itself and a Unix timestamp. Note that the example shown has its SHA-256\\
    %     string truncated.
    % }
    % \\
    %--------------------------------
    % \hline
    scans
    &
    \makecell[l]{%
        Dictionary of results for each security vendor, with the \texttt{detected} sub-field being true\\should that particular vendor flag a URL/submission as malicious.
    }
    &
    \makecell[l]{%
        Vendor-1: \{detected:false\}\\
        Vendor-2: \{detected:true\}\\
    }
    \\
    % \hline
  \end{tabular}
  \caption{Example of metadata within submissions}
  \label{tab:submission}
\end{table*}

\section{Background \& Related Work} \label{sec: background}

Colloquially known as links, URLs are simply pointers to content on the Internet \cite{rfc1738}. Rather than downloading and re-distributing an already online video, a person instead shares its URL with others. This eases the sharing of data, with URLs present in social media posts \cite{lee2013warningbird}, websites, emails and phone messages \cite{reaves2016detecting}. This same simplicity and ubiquity has led to abuse, with deceptive URLs that ensnare users \cite{quinkert2020bethephisher} (such as the second example in Table \ref{tab:urls}).

Breaking down a URL's components, the \textbf{scheme} for most URLs is \texttt{http} with increasing use of \texttt{https}. Attackers can use \texttt{https} to lull end-users into a false sense of security \cite{fbi2019cyber}. \textbf{Sub-domains} are entirely within the control of attackers, and is a favoured way to trick end-users as to the legitimacy of malicious URLs \cite{quinkert2020bethephisher}. \textbf{Domains} are purchased from authorised registrars depending on the \textbf{suffix}. For example, to purchase \texttt{mysite.net.au}, one must go thru a commercial reseller appointed by Australia's auDA \cite{auda2020about}. The \textbf{path} refers to the exact location of a page, file or other asset. They may contain \textbf{queries} that provide additional information to the server at the point of request, and includes any text after a ``?'' character.

% \begin{itemize}
%     \item The \textbf{scheme} for most URLs is \texttt{http}, with \texttt{https} use increasing for both legitimate and malicious purposes. As seen in the given example, attackers can use \texttt{https} to lull end-users into a false sense of security \cite{fbi2019cyber}.
%     \item \textbf{Sub-domains} are entirely within the control of attackers, and is a favoured way to trick end-users into believing the legitimacy of their malicious URLs \cite{quinkert2020bethephisher}.
%     \item \textbf{Domains} are purchased from authorised registrars depending on the \textbf{suffix}. For example, to purchase \texttt{mysite.net.au}, one must go thru a commercial reseller appointed by Australia's auDA \cite{auda2020about}.
%     \item The \textbf{path} refers to the exact location of a page, file, post, or other asset. They may contain \textbf{queries} that provide additional information to the server at the point of request, and includes any text after a ``?'' character.
% \end{itemize}

\subsection{VirusTotal Platform}

\begin{figure}[b]
  \centering
  \includegraphics[width=1.0\linewidth]{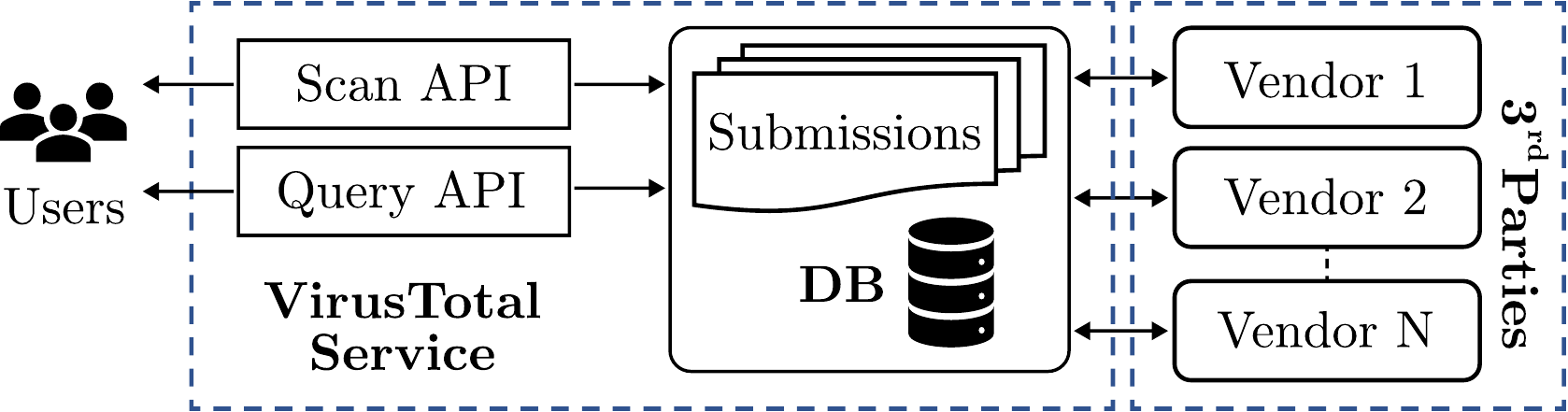}
  \caption{Users, vendors and the VirusTotal service}
\label{fig:virusTotalAPI}
\end{figure}

Given the known scale of the issue even as far back as 2011 \cite{ma2011learning}, multitudes of suspicious URLs are therefore sent hourly to cyber-security vendors such as PhishTank and Kaspersky. Each vendor employs their own customised engine to classify and thus detect malicious URLs, with varying levels of success \cite{peng2019opening}.

Building atop of such services, the VirusTotal platform can be thought of as an aggregator of results from a wide range of vendors \cite{song2016learning}. Essentially, it acts as a convenient front-end that allows researchers to query the status of suspicious URLs across a variety of scan engines available online. Figure \ref{fig:virusTotalAPI} depicts this relationship between users, security vendors and the VirusTotal platform.

While this offers much needed simplicity for the academic community, it must be noted that vendor labels do not always agree \cite{zhu2020benchmarking}. For example, a URL that has been flagged as malicious by one vendor may not necessarily share the same fate with another. Nonetheless, the platform has been extensively used by researchers to either label existing data or collect datasets for training and evaluation of algorithms. While past focus has been on both files \cite{cheng2018towards,kim2017certified, kim2018broken, korczynski2017capturing, lever2017lustrum, li2018jsgraph, szurdi2017email, wang2018beyond, wong2018tackling, xu2018vmhunt} as well as suspicious IP addresses and URLs \cite{catakoglu2016automatic,hong2018you, miramirkhani2018panning,oprea2018made, neupane2014neural,razaghpanah2018apps, sarabi2018characterizing,sharif2018predicting, tian2018needle,wang2014whowas,xu2014autoprobe, zuo2017smartgen}, the work herein aims to cluster and understand the dynamics of submitted URLs within the VirusTotal platform. That is, we are primarily concerned about characterising the URLs themselves via the metadata available for each submission, as opposed to the raw content that they point to.

\subsection{Vendor Labelling}

We note the slight contention within the community with regards to vendor labelling accuracy and counts within VirusTotal submissions. Most have defined a threshold, $t$, to classify submissions. That is, a URL is deemed malicious if the number of vendors flagging it surpasses said threshold. While most prior art \cite{catakoglu2016automatic,hong2018you,miramirkhani2018panning,razaghpanah2018apps,sarabi2018characterizing,tian2018needle,zuo2017smartgen} has cast wider nets by setting $t=1$, some have been more conservative by setting $t={2,3}$ \cite{oprea2018made,sharif2018predicting,wang2014whowas}, or even $t=5$ \cite{le2018urlnet}.

In contrast, the work herein is not directly concerned with labelling particular URLs as benign or malicious, and do not set arbitrary thresholds to filter submissions. Rather, we take submissions as a whole, cluster them by their metadata, and calculate the average vendor hits for each cluster. For example, if \texttt{bad.com} and \texttt{worse.com} are in a single cluster that points to the same malware, and are flagged by 4 and 2 vendors respectively, then their cluster would have a mean score of 3. Even then, we do not use these derived values to label or filter URLs. They are instead used to further characterise the clusters they represent and provide more insights into those campaigns that are secondarily verified as malicious.

\subsection{Dataset Description}
With kind assistance from VirusTotal representatives, we collected a large number of URL submissions via their \texttt{/url/feed} private API endpoint\footnote{https://developers.virustotal.com/reference\#url-feed}. These submissions stretched over a period of almost 2 months, from 2\textsuperscript{nd} Dec 2019 thru to 30\textsuperscript{th} Jan 2020. Submitted URLs were checked by an average of 71.9 third-party security vendors, with a majority of 94.18\% of submissions checked by 72 vendors.

Each submission stores the URL in question, along with various metadata around it, in multiple and nested fields in JSON format. In Table \ref{tab:submission}, we've described and shown examples of the most pertinent metadata. The primary field of concern here is the content hash (stored as \texttt{Response content SHA-256}), which is the computed hash of the \textit{content} that the URL points to. For example, if both \texttt{(bad.site/malware.exe)} and \texttt{(new.info/some.exe)} point to the same malicious executable, their submissions would have the same content hash. This is the straightforward mechanism with which we cluster submissions and arrive at our various findings.

\begin{table}[b]
\centering
    \begin{tabular}{r | l} 
    
    276,098,482 & Submissions with content hash \\
    35,489,645 & Submissions without content hash \\
    \textbf{311,588,127} & \textbf{Total submissions} \\[0.4em]
    
    52,458,809 & Unflagged unique URLs\\
    24,579,130 & Flagged unique URLs\\
    \textbf{77,037,939} & \textbf{Total unique URLs} \\[0.4em]

    %  Clusters with only one submission & 75,372,707 \\
    %  Unflagged clusters with 2+ submissions & 4,943,058 \\
    %  Flagged clusters with 2+ submissions & 2,617,572 \\
    %  \textbf{Total unique clusters} & \textbf{82,933,337} \\[0.3em]

    \end{tabular}
\caption{General details of collected dataset}
\label{tab:dataset}
\end{table}

Striving for completeness and accuracy, the work herein pre-processed, aggregated and analysed over 311M submissions encompassing 1.9TB of uncompressed text data. While studying a smaller subset (e.g., 1 week duration) would be easier, it would not have captured the full depth and width of malicious campaigns, and thus limited our findings. Table \ref{tab:dataset} highlights the general aspects as well as the scale of the data that was collected, specifically noting:

\begin{itemize}
    \item 11.39\% of the submissions (over 35.4M) had blank or entirely missing content hashes. These did not exhibit clear patterns and were uniformly spread, so it was not a case where data for a period was corrupted. Possibly, there were intermittent issues that prevented the recording of content hashes.

    \item There were only 77M unique URLs within the dataset, which is unsurprising as the same URL can be repeatedly submitted by multiple users. Thus, the naive mean rate of resubmissions is 4.044 over a 2 month period. Further, 24.5M URLs (31.9\%) were flagged by at least one vendor as malicious.

\end{itemize}

%% file: Sections/3_Individual_submission_char.tex
\section{Characterization of Submissions}
% \begin{comment} 
% Here we give stats about submissions in our dataset and the corresponding key findings, e.g., (Each bullet point represents a subsection throughout the skeleton on the paper.)
% $\bullet$ How many total submissions in a span of two months, submissions per day, ... (give stats in Finding 1 goes here)
% $\bullet$ How many benign/malicious submissions in our dataset? (this could be submissions with zero positives, etc. for benign case and for malicious case should have at least one positive) (Finding 2 goes here)
% \subsection{Temporal Characteristics} What is the rate of submissions, e.g., are they received at constant rates over a given period, or is there any 'storm' like events e.g., sudden increase in submissions.
% \subsection{Geographical distribution} of submitted submissions to VT. (Finding 3)
% \begin{figure}[!ht]
% \centering
% \includegraphics[width=1.0\linewidth]{Figures/distribution-positives.pdf}
% \caption{Distribution of Positives. \sharif{We could also show the bars with \% instead of \#. That may make them more representatives}}
% \label{fig:PositivesCount}
% \end{figure}
% \end{comment}

We begin by providing a statistical overview of the entire dataset, including general insights, temporal features, and the general performance of security vendors. It is hoped that the findings here would aid in the development of future detection strategies.

% Here we provide a statistical overview of the data we collected including general insights, temporal features, observed ``storms'' of sudden unexplained submissions, and the efficacy and detection time of security vendors within the VirusTotal (VT) platform.

% There is a need to look at the entire dataset from a high-level, and detail important overall characteristics that would aid future detection and defensive approaches by the community.

% In this section, we delve into a statistical overview of the dataset and highlight important operational insights of VT vendors. 
% We evaluate the performance of VT vendors in terms of the number of submissions with respect to detection. Next, we show temporal characteristic of the malicious URLs submitted to the VT and the time taken for detection by vendors. We delve into the statistical characteristics of daily submissions. We also discovered and characterized rare ``storms'' of sudden unexplained submissions, providing support that reports of high submission volumes by individuals do occur. Finally, we develop a method to rank VT vendors based on their detection capabilities. \newline

\subsection{General Findings}

\textbf{Finding 1:} \textit{Attacks seem rampant in the United States (US).} The US leads the way with a quarter of all of submissions (78,687,777 -- 25.25\%), followed by South Korea (47,973,096 -- 15.4\%),  Japan (38,578,771 -- 12.38\%), and Germany (32,681,855 -- 10.49\%). As the world's technological hub it is perhaps reasonable that most submissions originate from the US from businesses that use VirusTotal via its API. Also, with populations that are more technology literate, more people in such developed countries are aware of, and make use of, the manual submission features of VirusTotal.

\textbf{Finding 2:} \textit{Majority of submissions were automated, with a large percentage from a select few.} By grouping on the \texttt{interface} sub-field within each submission, we found an overwhelming 91.28\% (284.4M) of them came thru the \texttt{API} category, with \texttt{email}, \texttt{web}, \texttt{community} respectively scoring 7.33\% (22.8M), 1.06\% (3.3M) and 0.32\% (1M). This shows that most use of the platform is automated, perhaps from organisations that have commercial agreements with VirusTotal. Further, based on the \texttt{submitter\_id} field, we found that just two organisations were responsible for 30.42\% of all submissions, with both contributing $\sim$47M each. In fact, just ten organisations were responsible for close to half (48.34\%) of all submissions.

\textbf{Finding 3:} \textit{58.98\% of submissions were unflagged.} We observed that 183,785,016 submissions were not flagged at all by any of the vendors within VirusTotal. Going by the reasonable premise that only suspicious URLs are being submitted, this means that more than half of submissions were either truly benign or undetected malicious. In the first case, this perhaps signals the need for more work such as \cite{albakry2020what, quinkert2020bethephisher} to understand why users may deem benign URLs as suspicious. In the second case, this points to the need for more effective methods to better detect attacks as they occur.

\textbf{Finding 4:} \textit{17M unique pieces of content were flagged}. By grouping on submissions' content hashes, we identified that 20.6\% of unique hashes (17,116,732) within the dataset was potentially malicious. This was pulled from the corresponding 41.02\% of submissions (127,803,111) where at least one vendor has flagged their attached URLs as malicious. A majority of the 17M consisted of single submissions, where a unique hash had only one submission attached to it. As mentioned prior, there were 2.6M flagged clusters where the unique hash had 2 or more submissions attached.

Given the contention between the multitude of scan engines, we use the flagged status of submissions (and derived clusters) only as a signal for further analysis. We have erred on the side of caution and set a bare minimum threshold to guide our investigation. Simply, if a cluster has a single vendor flag within it, we consider it as potentially malicious, albeit at a lower risk than a cluster with multiple vendor hits. Only clusters with absolutely no hits at all amongst its submissions would be considered clean.

\subsection{Temporal Characteristics}

\begin{figure}[t]
  \centering
  \includegraphics[width=1.0\linewidth]{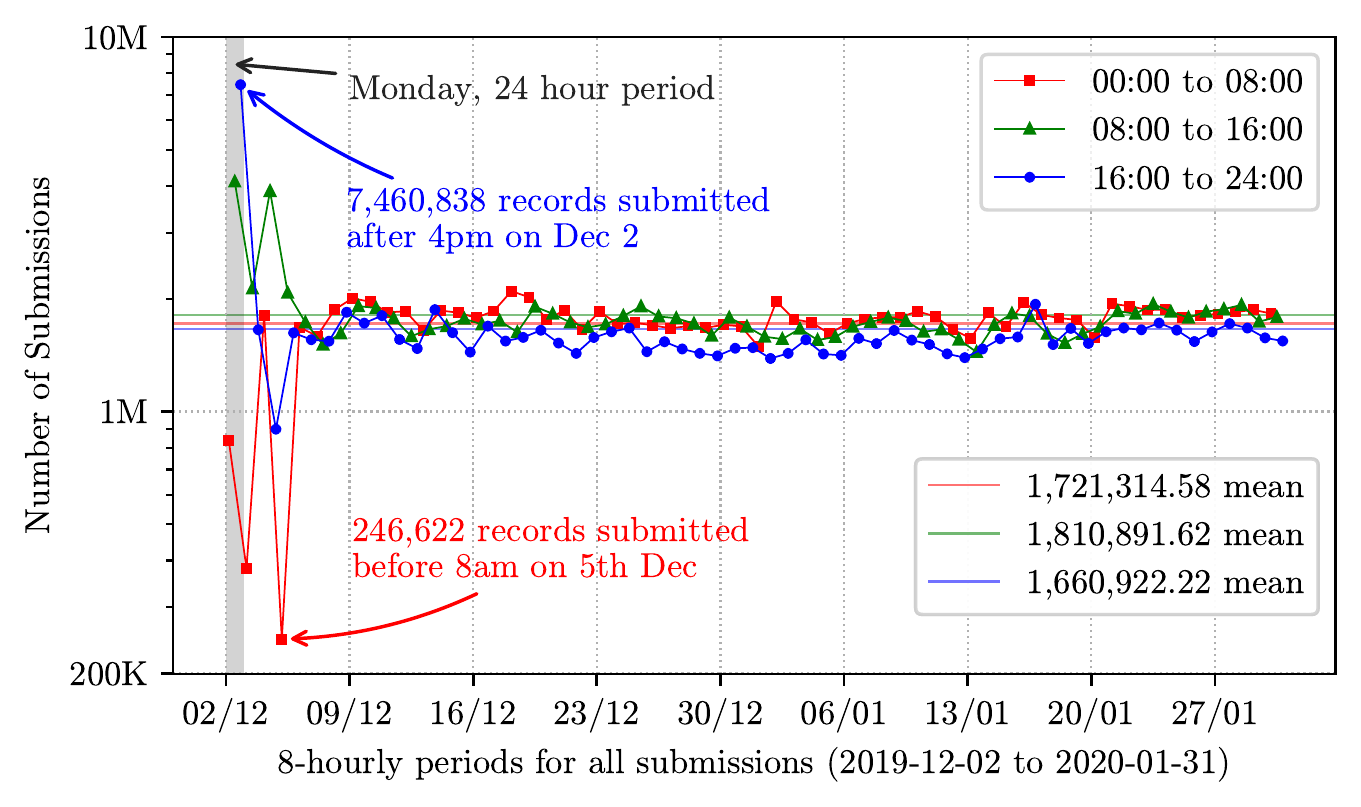}
  \caption{Scan times of all submissions}
%   \caption{In this plot of daily average reports, we observed a stationary reports submission volume distribution of clean and malicious URLs with a weekly periodicity during our study. Major events in our study are labelled.}
\label{fig:submissions over time}
\end{figure}

Daily scan volumes are shown in Figure~\ref{fig:submissions over time}, where we divided per day arrivals into three equal 8-hour chunks starting midnight, 8am and 4pm. As there was no timezone information within the dataset, it is assumed that all times were recorded as UTC. Outliers were observed in daily volumes from December 2\textsuperscript{nd} to 5\textsuperscript{th}, which we will detail later. Otherwise, daily submission counts remained stable.

\textbf{Finding 5:} \textit{We observed stationary daily submission volumes over a holiday period.} We fit a linear model to our weekly average submission volume after normalizing the data, finding a slope of -0.0290, indicating almost no change in the rate of submissions over the study period. We also fit a model after discarding the anomalous peaks from December 2-8, finding an even smaller slope of 0.0147.

This implies that the overall submission rate is fairly constant which is somewhat surprising given the time of year. It is reasonable to think that attackers would time their campaigns to coincide with key holiday events (e.g., schemes that purport to be Christmas sales). There would also be reasonable assumptions that submissions would drop as industries close and communications decrease. In any case, neither of these hypotheses materialised in the data.

% \textbf{Finding 4:} \textit{We observed a stationary daily submission volumes over a holiday period.}

\subsection{Outbreaks: High submission volume events}
% \ejaz{Mardi, please do not delete my edits just use `\%' to comment-out, and then add your edits.}

As mentioned, we observed ``outbreaks'' in the daily submission counts, where values were abnormally high (2\textsuperscript{nd} to 5\textsuperscript{th} Dec). To mathematically identify outbreaks, we used a uniform mechanism to compare submission volumes relatively across all submissions in the collected period. To this end, we used a z-score which is defined as $z = (x - \mu)/\sigma$, where x is a data point in the distribution, $\mu$ is the mean and $\sigma$ is the standard deviation. This was computed over the same 8-hourly periods for the entire 2 months. A z-score of 1 for a specific time period of a day indicates that the submission volume is a single standard deviation away from the mean. A conservative z-score of 2 was chosen as the limit to identify abnormally high submission volume events. Values greater than 2 indicated that the submissions within the specific time period were 2 standard deviations away and is an intuitive indication of outliers.

\textbf{Finding 6:} \textit{We observed one instance of an outbreak spread across three consecutive days.} During this outbreak, the average number of submissions spiked to 347,747 per hour compared to 210,270 submissions per hour under normal conditions, which is $\sim$40\% increase in submission volumes. This indicates that outbreaks are rare events, at least within the collected dataset. The term ``outbreak'' does not imply that the submissions reported were malicious as we cannot attribute a source for such events. However, there are reported instances of widespread phishing campaigns targeting numerous organizations across various industries. For example, on 2020-12-02, there was a wave of phishing campaigns targeting 28 organizations~\cite{mandiant_phishing_campaign}. Phishing emails from 26 unique email addresses and containing URLs with at least 24 unique domain names were used, all of which were engineered to lure victims to download malware. We surmise that such large-scaled phishing campaigns could result in abnormally high submissions rates.

\subsection{Vendor Effectiveness}

\begin{figure}[t]
  \centering
  \includegraphics[width=1.0\linewidth]{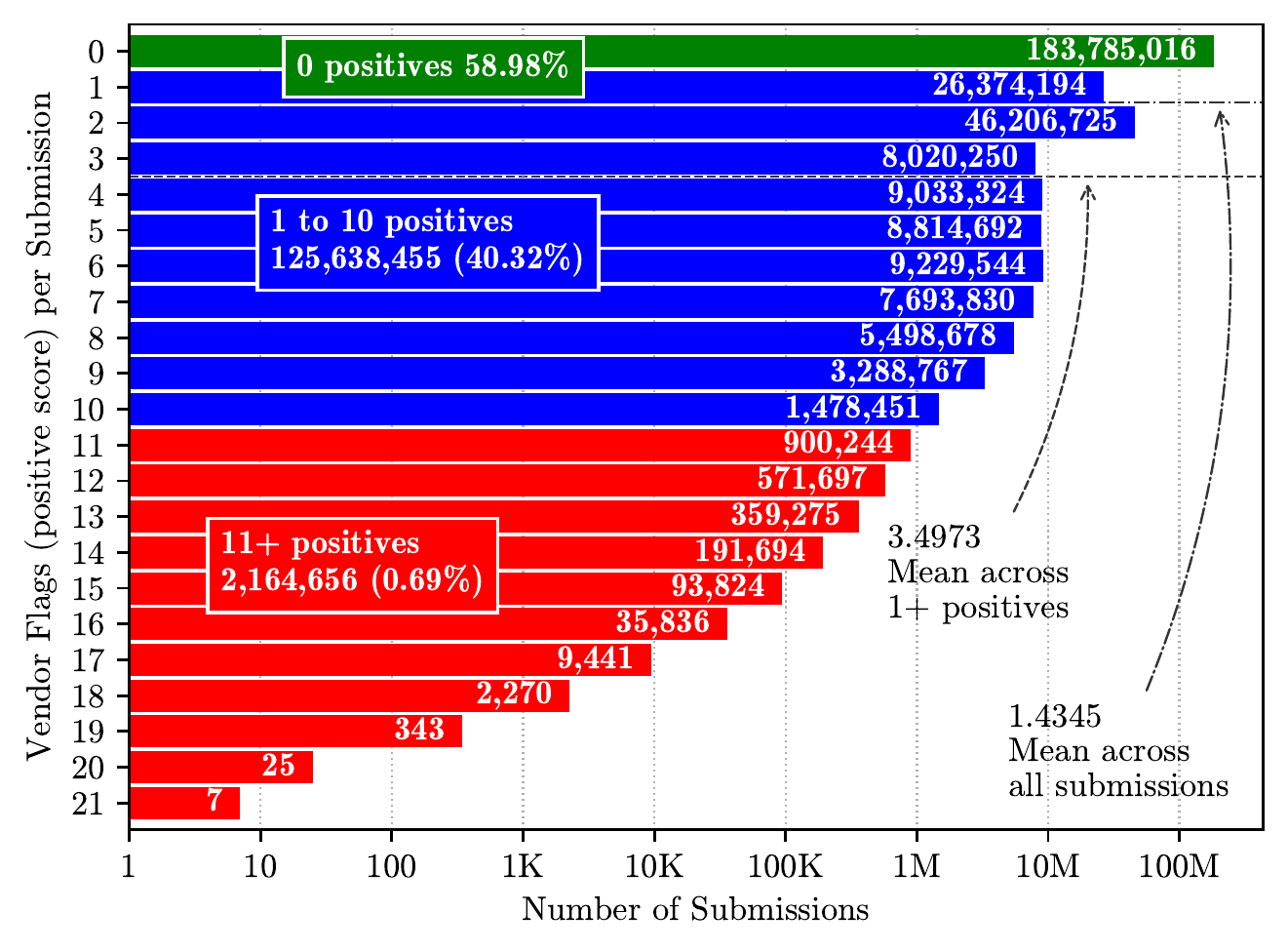}
  \caption{Spread of submissions by number of vendor flags}
\label{fig:positive distribution}
\end{figure}

Submitted URLs are analysed in parallel by a median of 72 security vendors to determine if they're benign or malicious. As seen in Table \ref{tab:submission}, each submission has a \texttt{positives} field that contains the number of vendors that have flagged it. In Figure~\ref{fig:positive distribution}, we delved deeper into the distribution of submissions with respect to vendor flags, and observed that detection performance varies significantly with most submissions being flagged by only a select few vendors.

% A URL submitted to VT is analysed in parallel by more than 80 security vendors to determine if the submitted URL is benign or malicious (phishing attempt). Here we show the distribution of the number of security vendors flagging submissions. In Figure~\ref{fig:positive distribution}, we delved deeper into the distribution of submissions with respect to number of security vendors. We observed that detection performance vary significantly among security vendors -- most submissions are flagged by few security vendors. \newline

\textbf{Finding 7:} \textit{We observed that around 98.27\% (125.6M) of all flagged submissions were detected by 10 or fewer vendors.} This indicates that vendor detection performance is highly skewed and only a few of them are effective. However, it must be noted that vendors have different specialisations, with some better designed to detect phishing and others customised to detect other malware. We observed that only seven submissions were detected by the observed maximum of 21 vendors. More specifically, only 1.69\% of all flagged submissions were detected by more than 10 vendors (2.1M out of 127.8M), which indicates that a select few were responsible for most detections.

\textbf{Finding 8:} \textit{We observed that on average around 4 security vendors detected most of the flagged submissions.} This implies that if a URL is flagged by at least four vendors, it is reasonable to conclude that it is malicious. This is important as it will assist system admins when selecting a threshold when labelling URLs based only on VT vendors. For example, if at least $n$ number of vendors has flagged a URL, it is likely malicious. If an overly-restrictive threshold (e.g., $n$=20) is set, it may result in significant number of missed detections. Thus, appropriate thresholds may be set based on  Figure~\ref{fig:positive distribution}.

\subsection{Time to $n$ Vendor Flags}

With each submission containing a \texttt{first\_seen} field, the question becomes: what is the mean time required for submissions to be flagged by $n$ number of vendors? That is, how quickly do vendors catch up in flagging URLs after it was first submitted.

\begin{figure}[b]
  \centering
  \includegraphics[width=1.0\linewidth]{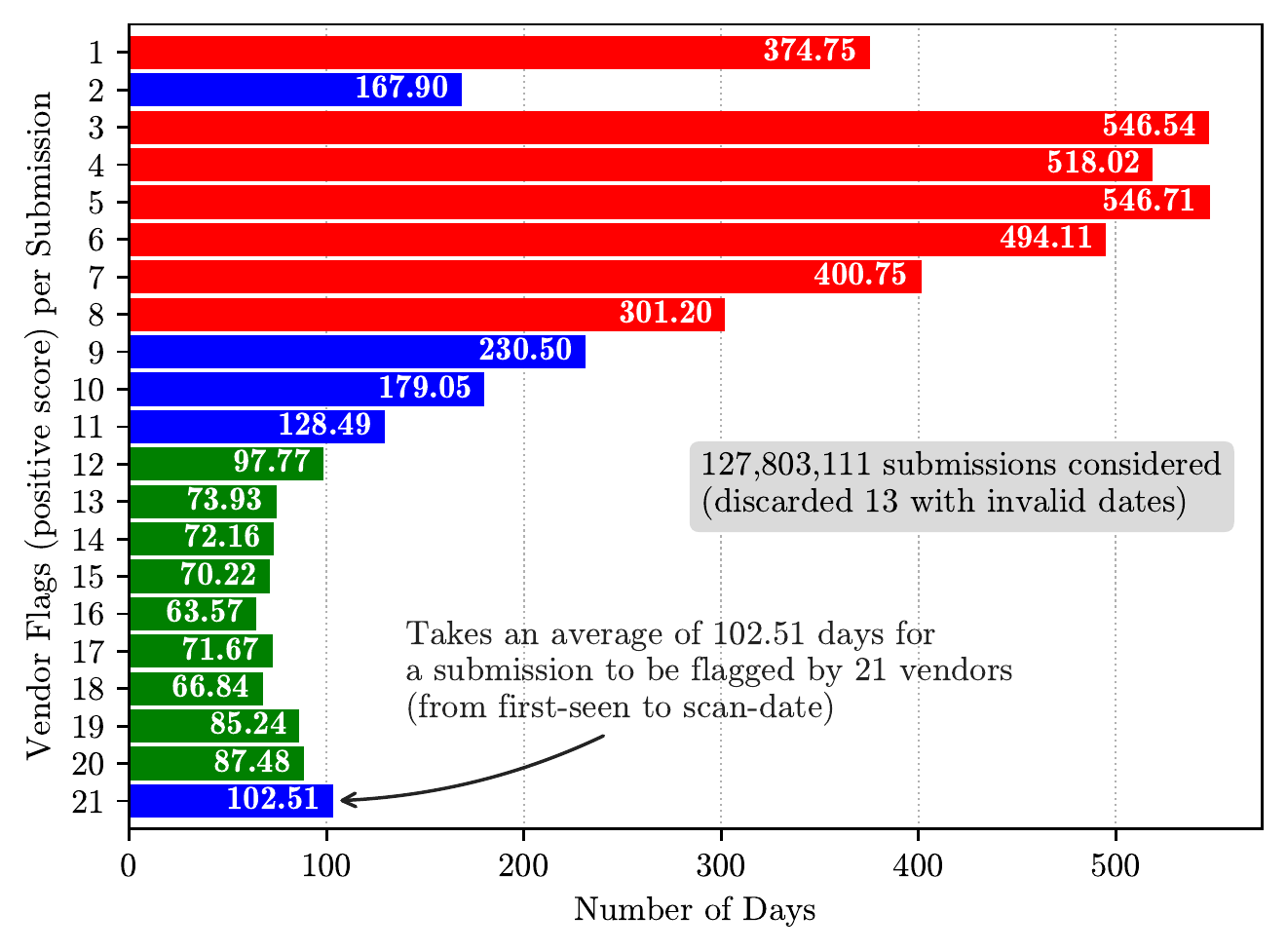}
  \caption{Mean time for submissions to reach $N$ flags}
\label{fig:positives first seen}
\end{figure}

\textbf{Finding 9:} \textit{The number of vendors flagging submissions is inversely correlated to the time these submissions were first reported.} More specifically, if a large number of vendors detected a particular set of submissions, it is more likely that these submissions are recently reported to the VT. Conversely, if submissions were flagged by only a few, it is are more likely that they were reported comparatively long ago. Intuitively, it is expected that more vendors would flag malicious URLs over time but surprisingly, this relationship between the time and vendors is quite opposite. Figure~\ref{fig:positives first seen} shows time-vendor distribution for the flagged submissions.

\textbf{Finding 10:} \textit{Submissions flagged by 12 or more vendors are detected significantly earlier compared to others.}
There could be two possible reasons for this. First, it indicates that some security vendors are trained on specific types of attacks and consequently, they are able to flag URLs in submissions promptly. We see this in Figure~\ref{fig:positives first seen} for the time distribution of 12 to 21 vendors. Second, vendors may rely on others when flagging URLs. For example, if 10 vendors flag a particular URL, others may also follow suit.

%% file: Sections/4_Campaign_Identification.tex
\section{Campaign Characteristics}

Here, we detail the clustering of submissions, and the subsequent identification and verification of confirmed malicious campaigns.

\subsection{Clustering}

% \begin{table}[b]
% \centering
%     \begin{tabular}{r | l} 

%     75,372,707 &
%     Clusters with only one submission \\
    
%     4,943,058 &
%     Unflagged clusters with 2+ submissions \\
    
%     2,617,572 &
%     Flagged clusters with 2+ submissions \\
    
%     \textbf{82,933,337} &
%     \textbf{Total unique clusters} \\[0.3em]

%     \end{tabular}
% \caption{Details on clusters within dataset}
% \label{tab:clusters_dataset}
% \end{table}

Each submission contains a hash code that is unique to the \textit{content} that its URL points to. This is stored within the 
\texttt{Response content SHA-256} field. Essentially, when a URL is given to VirusTotal, the file or HTML page that it refers to is retrieved, and a hashing function is run over it. It is important to note that the resultant output of the hash function is a unique identifier for the input content.

% Within each submission, we observed a \texttt{Response content SHA-256} field, which contains a hash code that is unique to the \textit{content} that the URL points to. Essentially, when a URL is given to VirusTotal, the file or HTML web-page that it refers to is retrieved, and a hashing function is run over it. It is important to note that the resultant output of the hash function is a unique identifier for the input content.

\begin{figure}[t]
  \centering
  \includegraphics[width=0.8\linewidth]{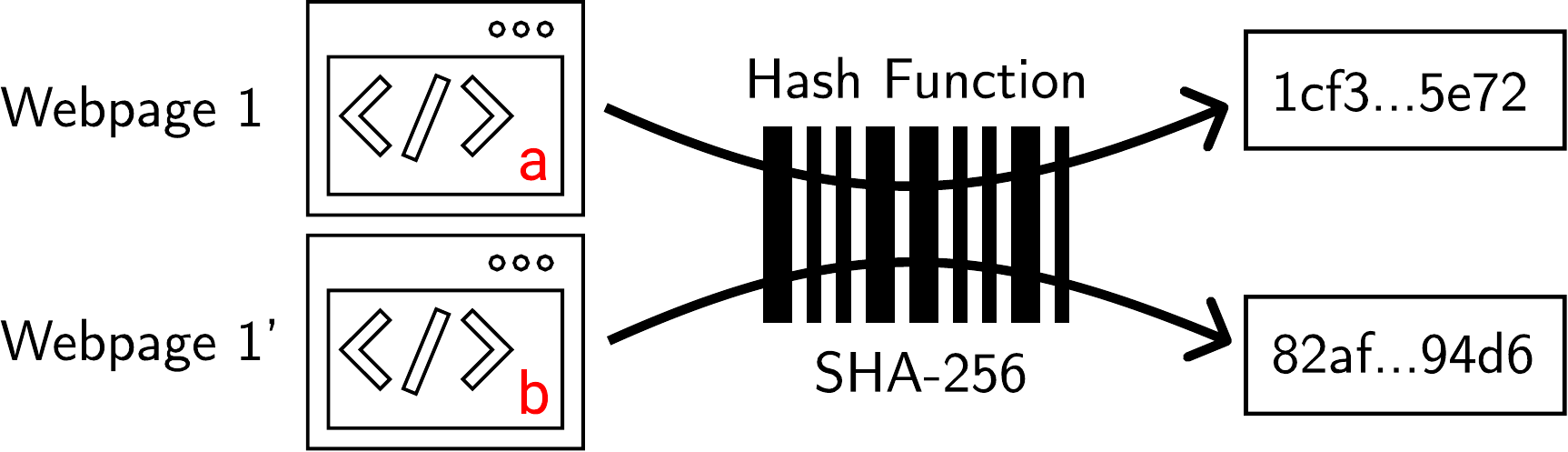}
  \caption{Showing how a single character change in a URL's content leads to a different hash code.}
\label{fig:hash}
\end{figure}

Figure \ref{fig:hash} illustrates this process, where \texttt{Webpage 1} is fed into the hash function and the resultant code is \texttt{1cf3...5e72}. However, if a single character \texttt{a} is changed to \texttt{b} as seen in \texttt{Webpage 1'}, the output hash code is then completely changed to \texttt{82af...94d6}. This is the mechanism with which we can cluster seemingly disparate submissions together. For example, if both \texttt{www.321.site/page.html} and \texttt{xyz.com/index.html} result in identical hash codes, we can assume that they point to the same content (possibly on different servers). Accordingly, if either is flagged, then both are malicious.

\begin{figure}[t]
  \centering
  \includegraphics[width=1.0\linewidth]{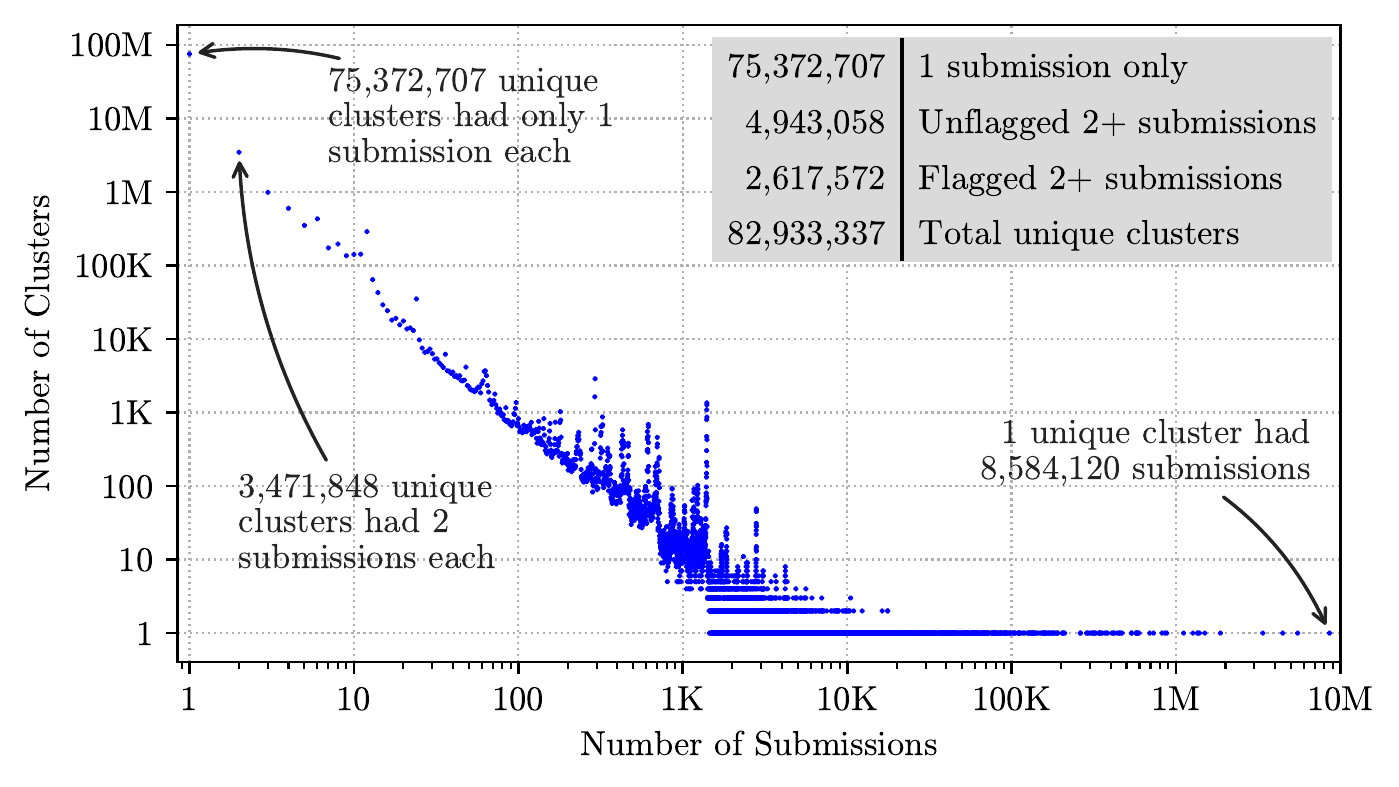}
  \caption{Distribution of clusters by number of submissions}
\label{fig:clusters_ditribution}
\end{figure}

This is the method with which we organise the 311M submissions into coherent clusters for further examination. Instead of trying to find similarities between the URL strings themselves in the hopes of finding groups, we simply rely on the reported content hashes within each submission's metadata. As noted previously in Table \ref{tab:dataset}, 11.39\% of all submissions did not have valid content hashes, which we ignore given the overwhelming number of remaining submissions. This results in the 82.9M clusters seen in Figure \ref{fig:clusters_ditribution}, where we see that a majority 75.3M had only one submission attached. That is, given a particular submission $S$ with a unique content hash $H$, then $S_H=1$. We've ignored these for now, given time constraints and our stated focus on campaigns of malicious URLs.

Figure \ref{fig:clusters_ditribution} shows the distribution of all clusters according to their number of submissions, where we see the single-submission cluster on the extreme left. There was one cluster containing 8.5M submissions (extreme right) which we theorise could be related to a standard \texttt{404} error page. There were 7.5M clusters with at least 2 records ($S_H >= 2$), of which 2.6M (34.6\%) had at least one submission flagged by at least one vendor as malicious. This strongly signals the presence of concerted campaigns, especially given the dataset's relatively short time frame of 2 months. These are the clusters we focus on as they are likely to contain malicious URLs.

% \newcolumntype{L}[1]{>{\centering\arraybackslash}m{#1}}

\begin{table*}[t]
% \footnotesize
\centering
  
 \begin{tabular}{%
    L{0.074\linewidth}%
    L{0.05\linewidth}%
    L{0.05\linewidth}%
    L{0.06\linewidth}%
    p{0.4\linewidth}%
    p{0.2\linewidth}%
 }

    Content Hash &
    URLs &
    Submis-sions &
    Mean Positive &
    URLs Detected (GSB Secondary Verification) &
    Example Undetected URLs \\

    \hline
 
    % 0911-f5a3 &
    % 1 &
    % 205 &
    % 0.004878 &
    % book.dypics[.]com/search/\%E3\%80\%90Survi... &
    % N/A \\
    
    5db9-1212 &
    3 &
    3,426 &
    0.005254 &
    4ksudckusdkc[.]space &
    % alwkdlka[.]club, qoiwdjoiqjw[.]club \\
    alwkdlka[.]club \\
 
    % 2e1c-b3b5 &
    % 4,282 &
    % 4,647 &
    % 0.013988 &
    % www.ultrafiles[.]net/AfYka &
    % bit[.]ly/2P8HlTr, theseblogs[.]com/C8QVm\\
    
    d992-844c &
    1,053 &
    596,116 &
    0.033406 &
    firsttime-traffic41[.]loan &
    % zap385438-4.plesk05.zap-webspace[.]com\\
    manerck[.]com\\

    % 32b4-9892 &
    % 179 &
    % 949 &
    % 0.033720 &
    % googieapls[.]com &
    % fonts.proxy.ustclug[.]org, fonts.lug.ustc.edu[.]cn\\
    
    16f3-21a6 &
    2 &
    12 &
    0.083333 &
    everyday-vouchers[.]com, sweepstakehunter[.]com &
    N/A \\
    
    % 1e3b-83df &
    % 251 &
    % 263,114 &
    % 0.046345 &
    % znp.fy.h.nipqa.arg.r.de.a2ip[.]ru &
    % download.amd.com/Desktop/aod\_setup\_4.3.1.exe\\
 
\end{tabular}

\caption{Three (out of 974) low-risk flagged clusters that were confirmed as malicious campaigns}
\label{tab:low-risk-examples}
\end{table*}

\subsection{Vendor Labelling}

To help guide our analysis, we utilize the preexisting vendor labels within each cluster. Recall that each submission contains a list of security vendors that have labelled that submission's URL as benign or malicious. Most vendors claim to use some form of machine learning, which despite their efficacy, is known for False-positive and False-negative behaviours. Besides, the pre-existing labels for each URL were issued simultaneously at the point of submission, and there is an opportunity for us revisit their labelling with knowledge from the present day. Therefore, we implemented secondary labelling to act as an expanded ground truth for our analysis.

To that end, we examined three candidate phishing block-lists in wide use today: Google Safe Browsing (GSB) \cite{GSB}, OpenPhish (OP) \cite{OP}, and PhishTank (PT) \cite{PT}. According to the recent empirical analysis done in \cite{bell2020analysis}, GSB is by far the largest block-list available, with over 17 times as many URLs compared to PT and OP combined. It is also used by a majority of web browsers such as Chrome, Safari and Firefox and protects roughly four billion devices per day by showing warnings to users should they attempt to access malicious URLs. This provides GSB with the largest pool of users (over 600M) with which to obtain frequent reporting \cite{bell2020analysis}. Hence, we selected GSB as the secondary ground truth labelling mechanism.

\subsection{Identification and Verification}

\begin{figure}[b]
  \centering
  \includegraphics[width=1.0\linewidth]{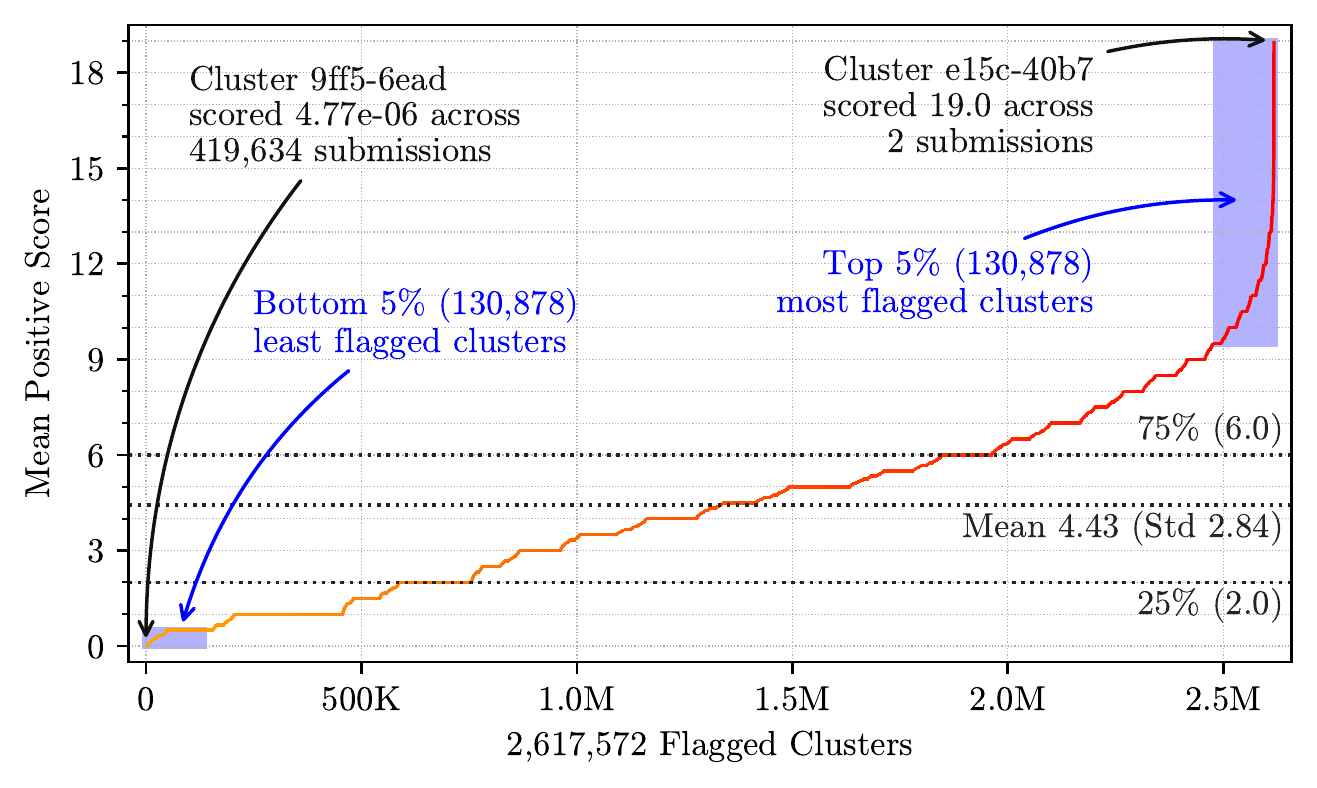}
  \caption{2.6M flagged clusters sorted by risk factor.}
\label{fig:clusters-risk}
\end{figure}

Thus, we turn our attention to the 2,617,572 flagged clusters that recorded at least one vendor hit within their submissions. These proved fertile ground for the verification of malicious campaigns via secondary GSB checks. Figure \ref{fig:clusters-risk} shows these clusters sorted from left to right according to a risk factor defined as follows:

\begin{itemize}
    \item \textbf{Positives}: Each submission contains a \texttt{positives} field which indicates the number of vendor flags against it.
    \item \textbf{Mean Positive Score}: This is simply the sum of all \textbf{Positives} divided by the submission count of each cluster.
    \item \textbf{Unflagged Clusters}: Clusters are deemed unflagged if they have mean positive scores of 0.0. That is, not a single vendor flag in any of the submissions within a cluster.
    \item \textbf{Flagged Clusters}: Conversely, this is when clusters have \textit{at least one} vendor hit (i.e., mean positive score > 0.0).
    \item \textbf{Malicious Campaigns}: Flagged clusters that are then verified by GSB as containing \textit{at least one} malicious URL.
\end{itemize}

For example, if a cluster had three submissions with positive scores of 7, 7, 8 respectively, then its mean positive score would be 7.33. Conversely, if a cluster had only a single positive vendor hit amongst 10K submissions, then it's score would be 0.0001. Subsequently, if either of these contain a URL that appears in the GSB block-list, then they are deemed malicious campaigns.

\textbf{Finding 11}: \textit{Flagged clusters with seemingly very low risk can contain verified malicious campaigns.} Given their low mean positive scores, one would expect that the bottom 5\% quantile of flagged clusters to perhaps be clear of malicious URLs. These 130,878 supposedly low-risk clusters contained 7,002,793 unique URLs within them and can be seen on the extreme left of Figure \ref{fig:clusters-risk}. However, when we reran their URLs against GSB, we found hits for 974 of the clusters. The confirmation of such clusters as malicious is concerning, given their comparatively low mean positive scores that range from as low as 0.001547 to 0.5 (with a mean of 0.3540).

This signals the need for better detection mechanisms, and that vigilance is required even when a cluster seems of very low risk. Of course, the numbers may be skewed in the sense that a wave of submissions with the same URL (and hence content) could dilute the mean positive score and thus the perceived riskiness of a cluster. We see examples of this with cluster 5db9-1212 in Table \ref{tab:low-risk-examples}, where just three URLs were repeatedly submitted and thus dragged down the mean positive score. In any case, this points to the fact that, taking all the vendors as a whole, even a single hit in a cluster is cause for concern. This is true especially for cluster 16f3-21a6, where there was only \textit{a single vendor flag} across all 12 of its submissions.

We also see the first evidence of large campaigns with cluster d992-844c. It contained seemingly unique URLs that all point to the same content, of which only one was confirmed as malicious by GSB. Whilst the sheer number of submissions seemed to dilute the apparent risk, it was no less dangerous given the diversity of URLs within it. There were 364 unique sub-domains, 782 unique domains, and 109 unique suffixes used throughout its 1,053 URLs.

\textbf{Finding 12}: \textit{GSB did not detect all malicious URLs within campaigns.} This is exemplified in cluster d992-844c in Table \ref{tab:low-risk-examples} where only a single URL was detected out of 1053 URLs. From the 974 discovered malicious campaigns, 939 had at least two unique URLs and the mean percentage of \textit{undetected} URLs within them stands at 61.07\%. That is, a majority of malicious URLs constantly escape detection, especially as the number of unique URLs increases. As context, the average number of URLs within these 939 campaigns stands at 2326.12, with the largest campaigns maxing out at 729,858.

\textbf{Finding 13}: \textit{94.5\% of unique URLs went undetected by GSB in high-risk flagged clusters.} On the opposite end in Figure \ref{fig:clusters-risk}, we verified all the unique URLs within the top 130,878 flagged clusters (i.e., the top 5\% in terms of risk). Given their high mean positive scores, it can be reasonably expected that the majority of URLs would indeed be malicious. However, GSB only detected 5.4\% of the 333,728 unique URLs contained within these high-risk clusters.

The argument here could be that these malicious campaigns were swiftly detected, dealt with by authorities and/or their instigators have since ceased their concerted attacks. This then may have prompted subsequent removal of the URL entries from GSB block-lists and thus bypassed our secondary verification. However, we posit that GSB would not be so quick to remove such entries, given the relative recency of the events. It has only been 14 months since data collection in Jan 2020 and re-verification in Apr 2021. There is also the fact that 17,895 malicious URLs (5.4\%) from the same clusters were indeed found in GSB block-lists.

\begin{table}[b]
% \footnotesize
\centering
  
 \begin{tabular}{%
    r|%
    l%
 }
    
    59,450 &
    Campaigns with a single unique URL \\
    18,360 &
    Campaigns with multiple unique URLs \\
    \textbf{77,810} &
    \textbf{Total Campaigns} \\[0.4em]

    59,450 &
    URLs in Single-URL Campaigns \\
    9,900,146 &
    URLs in Multi-URL Campaigns \\
    \textbf{9,959,596} &
    \textbf{Total Unique URLs} \\[0.4em]

    276,801 &
    Submissions in Single-URL Campaigns \\
    37,880,096 &
    Submissions in Multi-URL Campaigns \\
    \textbf{38,156,897} &
    \textbf{Total Submissions} \\[0.4em]

\end{tabular}

\caption{Characteristics of found malicious campaigns}
\label{tab:campaigns}
\end{table}

\textbf{Finding 14}: \textit{11,622 confirmed malicious campaigns were found in high-risk flagged clusters.} Despite the low URL detection rate above, we found a much higher incidence of malicious campaigns. This is perhaps to be expected given the high mean positive scores, which implies more vendors were flagging the URLs within each cluster and thus, a much higher likelihood of malicious campaigns.

On the other hand, only confirming 8.88\% of 130,878 clusters is seemingly low for a state-of-the-art block-list such as GSB. On average, the mean positive score for clusters detected by GSB was 11.36. Using this as a benchmark, there were 36,678 other high-risk clusters with higher mean positive scores that were missed by GSB. This relatively high number of undetected clusters is in line with the high number of undetected URLs seen earlier, and alludes to the known shortcomings of the block-list approach. In other words, should an attacker choose to rerun some of these campaigns today, the majority of web-browsers that rely on GSB would not be able to prevent their users from accessing these known bad URLs.

Finally, we do notice a significant drop in the number of unique URLs in the low-risk clusters (7M) to just 333,728 within these high-risk ones. An explanation could be that the campaigns here are being detected at a higher rate by vendors, and thus the URLs and domains that attackers use are shut down faster. This leads to less reporting as the instances of abuse wither and attackers give up on trying to use different URLs for the same malicious content.

% Here we discuss how we labeled.
% Labeling and sorting according to risk.
% Now we need to double validate against GSB!

% \subsubsection{Check Top Quartile Hashes}
% Check most flagged hashes to confirm our campaign detection approach. Is there any URL \textbf{not} detected for hashes.

% Example: Hash 1 has 2 unique URLs. One is detected by 19 vendors, but the other does not have any vendor flagged.

% \subsubsection{Check Bottom Quartile Hashes}
% To see if there are any hits from GSB or Quad9. (there are least flagged so we're trying to find URLs that have escaped detection).

% \subsection{VT Detection} Here we give our findings about VT vendors labeling approach. And provide findings what could be reasonable threshold based on the dataset. In our experience, even with positives equal 2 submissions are malicious. This should be highlighted here. (Finding 8, etc....)

% \begin{comment}
% \begin{figure}[!ht]
% \centering
% \includegraphics[width=1.08\linewidth]{Figures/distribution-hashes.pdf}
% \caption{Distribution of clusters within the dataset based on Collided SHA-256 \sharif{We may need to remove the numbers on the top right and keep the story simple that we found clusters.}}
% \label{fig:PositivesCount}
% \end{figure}
% \end{comment}

\subsection{Malicious Campaigns}

Concluding our initial analysis on the highest and lowest risk flagged clusters, we proceeded to run our secondary verification on the entire set of 2.6M flagged clusters. Our goal was to perform analysis on \textit{verified} malicious campaigns. The hope from the outset was that we would perhaps gather 100K campaigns from the top 5\% flagged clusters alone. However, the unexpectedly low numbers seen in prior sections subsequently forced us to widen our net.

\textbf{Finding 15}: \textit{There are 77,810 confirmed malicious campaigns within the dataset.} We successfully verified that 2.97\% of the 2.6M flagged clusters were indeed malicious. These campaigns have \textit{at least one} URL within them that has been marked by GSB as pointing to either deceptive or phishing pages (\texttt{SOCIAL\_ENGINEERING}), or malicious payloads (\texttt{MALWARE}), or even both. As all of a campaign's submissions are labelled with the same content hash, a single marked URL therefore implies that \textit{all} URLs within that campaign point to the same malicious content.

The number of confirmed campaigns seems low at only 77,810, which perhaps points to the limitations of the current block-list approach by GSB. On average, the mean positive score for these clusters verified by GSB stands at 6.2, with a standard deviation of 3.05. If we were to use this as a threshold, it is noted that 605,323 other flagged clusters with higher mean positive scores were not detected by GSB. That is, assuming that all flagged clusters were indeed malicious, GSB may have missed this large number of flagged clusters which comprise of 2,034,777 unique malicious URLs.

% Ideally, we would have preferred a higher number of confirmed campaigns, as the obtained population size seems low at only 77.2K. Nonetheless, we did obtain several useful insights from this small amount of campaigns.

% Furthermore, 

% While the number of confirmed campaigns may seem low at only 77.2K, we did obtain several useful insights.

\textbf{Finding 16}: \textit{The vast majority of malicious URLs come from campaigns that employ multiple unique URLs.} Table \ref{tab:campaigns} details the general characteristics of the confirmed campaigns, which we have broadly categorised into campaigns that have either a single unique URL, or multiple unique URLs. A majority (76.4\%) of the found campaigns consists of single URLs (i.e., the same URL is repeatedly submitted) but these only contributed 0.6\% (59,450) of unique URLs and 0.73\% (276,801) of submissions. The 18,360 multi-URL campaigns on the other hand, were responsible for the majority of URLs and submissions. This lends credence to our hypothesis that masses of URLs should be treated as coherent campaigns.

\begin{figure}[t]
\centering
\includegraphics[width=1.0\linewidth]{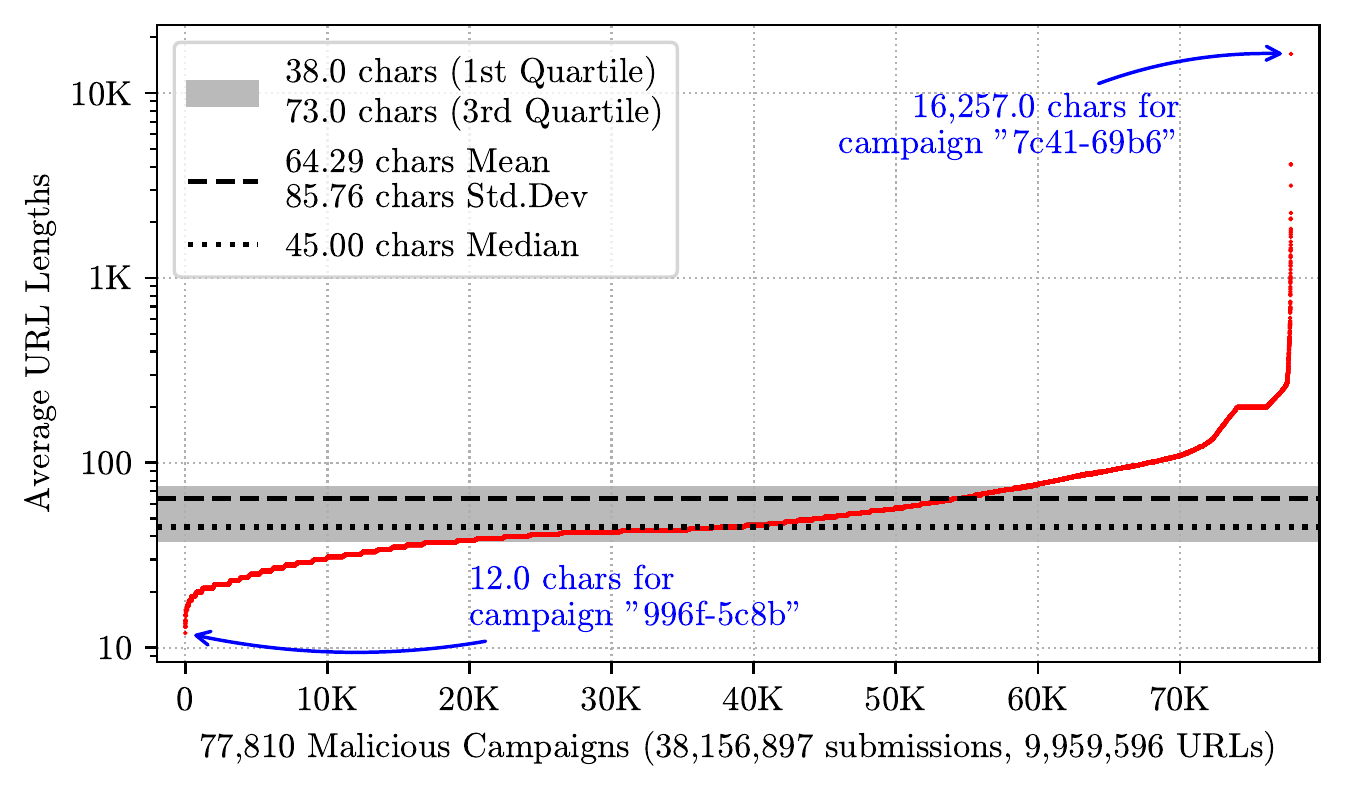}
\caption{Mean URL lengths of 77,810 campaigns}
\label{fig:url-lengths-campaigns}
\end{figure}

\textbf{Finding 17}: \textit{Campaigns favor the use of longer URLs.} The average URL lengths across campaigns (i.e., mean of means) stands at 64.29 characters. There is a standard deviation of 85.76 indicating a high degree of variance which we see in Figure \ref{fig:url-lengths-campaigns} with some campaigns on the extreme right employing URLs with tens of thousands of characters. 50.89\% of campaigns have average URL lengths of between 38.0 (1st quartile) and 73.0 (3rd quartile). This is likely due to attackers relying on increasingly longer URLs to deceive both end-users and commercial detection systems. As end-users are increasingly trained to detect malicious URLs, they grow to distrust shorter URLs that point to unknown locations \cite{albakry2020what} or comparatively short URLs with added terms in their domain names \cite{quinkert2020bethephisher}.

% The column width is: \the\columnwidth

% \hl{mardi stopped here}

% $\bullet$ Here we provide stats about number of campaigns detected. 

% $\bullet$ Ratio of submissions from campaign to total submissions.

% $\bullet$ Number of unique URLs/domains/IP/resolutions, etc. in campaigns. 

% $\bullet$ We can plot a graph where X-axis show \# of unique URLs in a campaign and Y-axis show number of campaigns, etc.

% \textbf{Finding XX:} \textit{We observed that most of the campaigns had 100 or less URLs -- out of 15,505 campaigns, only 1,040 campaigns had more than 100 URLs while 14,465 campaigns conduct phishing attempts using 100 or less URLs.}

\subsection{Campaign Metrics}

To systematically evaluate the characteristics of campaigns, we've defined and calculated metrics to quantify their behaviors:

% To systematically evaluate operational characteristics of campaigns, we define and calculate metrics to quantify the behavior of phishing campaigns. 

\noindent$\bullet$ \textbf{Campaign size}: Number of URLs deployed in each campaign.

% \noindent$\bullet$ \textbf{Campaign size}: Number of URLs deployed in each campaign, where a campaign is represented by a cluster of submissions having same hash.

\noindent$\bullet$ \textbf{Source distribution}: Ratio of the campaign size to the number of submission count reported by users in a campaign. A 100\% source distribution indicates that the campaign used a different URL for every submission reported at VT. This metric quantifies the rate at which campaigns attempt to created URLs. This metric tells us how diverse a campaign is in terms of the number of URLs created by attackers and correspondingly how many times they are reported.

\noindent$\bullet$ \textbf{Campaign footprint}: For footprint, we consider campaign metrics from both attacker (count of unique URLs deployed by a campaign) and victim (number of times these URLs reported by users) sides. We use the following expression to compute the footprint of a campaign: $fp = \mu U + (1-\mu)S$, where $U$ represents the count of number of URLs in a campaign, $S$ denotes the number of time these URLs are reported VT by the users. $\mu$ represents the weight assigned to each metric in the expression. We set equal weights(i.e., 0.5) for both the metrics in a campaign. The parameter $\mu$ can be tuned in order to prioritize certain metrics.

\noindent$\bullet$ \textbf{Domain/sub-domain diversity}: Ratio of the count of unique sub-domains in a campaign to the number of URLs in a campaign. A 100\% sub-domain diversity indicates that every domain in a set of URLs from this campaign was different. This metric is important for us to understand if a campaign is targeting specific global brands or tends to distribute phishing attempts across a wide range of users. In this case, lower sub-domain diversity can suggest a highly targeted campaign toward organizations.

% \noindent$\bullet$ \textbf{Path diversity}: As mentioned in Section~\ref{sec: background}, path of a URL refers to the exact location of an asset, Path diversity here refers to the unique number of paths in a campaign to number of URLs in a campaign. A 100\% path diversity indicates that every path in a set of URLs from a campaign was different. This metric indicates if an attacker changes the directory location of the malicious files every time he creates a new URL. 

\noindent$\bullet$ \textbf{Detection}: For phishing campaign detection, we evaluate detection rate using further two metrics: GSB and VT's security vendors. A 100\% GSB detection against a campaign indicates that all URLs in a campaign are flagged by GSB. However, for security vendors, we report the mean number of vendors who flagged URLs in a campaign. This metric helps us to understand the detection rate of URLs in a campaign against GSB and overall VT vendors.

% \begin{figure}[!ht]
% \centering
% \includegraphics[width=1.0\linewidth]{Figures/num-urls-cdf.pdf}
% \caption{Cumulative distribution function for number of unique URLs in 15,505 malicious campaigns}
% \label{fig:num_urls_cdf}
% \end{figure}

% Mean [Source distribution: 61.575309900032245, Footprint: 1369.6183489197033, Domain diversity: 66.57904611415672, path diversity: 68.09359238955176, mean positives: 5.997291196388262, gsb detection: 70.22657271847791]
% Total campaigns: 15,505
% Campaigns > 1 and < 100 URLs: 14,465 [63.41714621500173, 52.94434842723816, 68.46725820947114, 22.296796405115796, 69.17169789146214, 6.110197027307294, 74.32121050812306]
% Campaigns > 1000 URLs: 1,040   [35.957846153846155, 19682.78125, 40.31655769230769, 9.778740384615384, 53.09859615384615, 4.426923076923077, 13.275673076923077]

% \subsection{Collective detection} Here we give stats about our findings relating to collective detection, e.g., we take a cluster and label it. How VT label all the submissions in that cluster? Finding come here relating to collective detection vs individual labeling from VT.

After we defined several metrics, we compute them for each of the 15,505 phishing campaigns. Here, we interpret the metrics to comprehend how these campaigns differ from each other.

\textbf{Finding 18:} \textit{Phishing campaigns had an average source distribution of 61.58\%, which means that most campaigns attempt to deploy different URLs in each campaign.} We observed that 6,160 phishing campaigns with on average campaign size of 146 URLs (footprint of 160) had source distribution of 98.75\%. Moreover, 5,498 campaigns had source distribution of 100\%. The high source distribution rate shows that the campaigns are likely to create new phishing URLs frequently. 
The findings regarding the source distribution indicate that well-known blacklisting techniques that leverage whitelists or blacklists will not effectively detect URLs from many campaigns.  

We also observed that for 1,040 large-scaled campaigns (campaign size greater than 100 URLs), the source distribution reduced to 35.95\%. 
For such campaigns, the average campaign size, footprint, and number of reported submissions were 5,132, 19,682,and 34,233, respectively. 
Moreover, we observed that the largest phishing campaign with a campaign size of 1,391,080 URLs (footprint 4,987,600 and reported submissions were 8,584,120) has a source distribution of 16.21\%. Our findings indicate that the source distribution of a campaign decreases with increase in campaign size. This behavior could be attributed to the fact that as number of URLs increase, the corresponding reports from users increases non-linearly.
These statistics indicates that large-scaled phishing campaigns might use domain generation algorithms (DGAs) to generate a large number of domain names that could potentially be used as rendezvous points with their command and control servers. The large number of potential rendezvous points makes it difficult for security vendors to detect and take down the URLs from phishing such campaigns since the infected computers will attempt to contact some of these domain names every day to receive updates or commands. 

\textbf{Finding 19:} \textit{Phishing campaigns had an average domain and sub-domain diversity of 66.57\% and 21.45\%, respectively, with specific targeted organizations.}
We observed higher diversity in domain names compared to that of sub-domains. This is because attackers need to register the domain names following certain guidelines whereas for sub-domains, the users can choose names of their choice. Due to this flexibility, campaigns targeting particular organizations can use sub-domain names specific to the target organization, such as in targeted phishing. We observed that 153 campaigns had a domain and sub-domain had diversities of 2.29\% and 1.61\%, respectively, which indicates that attackers used a fixed set of domain and sub-domain names targeted towards few global brands and financial institutions. For instance, we discovered a large-scaled campaign with campaign size of 4,081 unique URLs that used only 12 domains and nine sub-domains targeting Apple brand (see Section~\ref{ssec: case study impersonation}).

\textbf{Finding 20:} \textit{We discovered that GSB performs well in detecting campaigns (having size 100 or less) achieving 74.32\% detection rate. However, GSB's detection performance drops to 13.27\% for relatively larger campaigns (size above 100).}
We observed that the performance of GSB's degrades with campaign size. For example, the overall detection rate for larger campaigns, i.e., campaign sizes greater than 10,000 URLs, the detection rate of GSB further drops to 2.46\%. 
One reason for this detection drop could be attributed to the fact that malicious URLs are often short-lived. In \cite{bell2020analysis}, it was reported that fewer URLs were residing in three popular blacklists (GSB, OP, and PT) as the time since they were blacklisted increases, which indicates that URLs are often short-lived. Moreover, none of the three blacklists enforce permanent-resident policy for URLs. They also observed that a significant number of URLs reappear once they were removed from blacklist. The lower detection rate for larger campaigns seems to be in consistent with the findings in \cite{bell2020analysis}.

% \noindent\textbf{Finding 19:} \textit{We discovered that GSB performs well in detecting campaigns (having size 100 or less) achieving 74.32\% detection rate. However, GSB's detection performance drops to 13.27\% for relatively larger campaigns (size above 100).}
% We observed that the performance of GSB's degrades with campaign size. For example, the overall detection rate for larger campaigns, i.e., campaign sizes greater than 10,000 URLs, the detection rate of GSB further drops to 2.46\%. 
% One reason for this detection drop could be attributed to the fact that phishing URLs are often short-lived~\cite{bell2020analysis}. 
% Bell et al. \cite{bell2020analysis} reported that fewer URLs were residing in three popular blacklists (GSB, OP, and PT) as the time since they were blacklisted increases -- indicating that phishing URLs are often short-lived. Moreover, none of the three blacklists enforce permanent-resident policy for phishing URLs. They also observed that a significant number of URLs reappear once they were removed from blacklist. The lower detection rate for larger campaigns seems to be in consistent with the findings in \cite{bell2020analysis}.

Like GSB, security vendors in VT also struggle to detect URLs as the campaign size grows. We observed that on average six security vendors detect phishing campaigns (size of 100 or below) and four security vendors detect large campaigns (size above 100).

%% file: Sections/5_Case_studies.tex
\section{Case Studies} \label{sec: case studies}

Here we provide examples of the attack strategies seen in campaigns by randomly selecting three campaigns and analysing them in terms of their scale, diversity, and detection by security vendors.

% This allowed us to label these campaigns and also helped us to selectively delve into campaigns with unique characteristics. We discovered that some of them use potentially misleading strategies to encourage victims to act.

% To get an overview of a phishing campaign, we randomly selected three campaigns and analysed them in terms of scale, diversity, and detection by security vendors. This allowed us to label these campaigns and also helped us to selectively delve into campaigns with unique characteristics. We discovered that some of them use potentially misleading strategies to encourage victims to act.

\subsection{Targeted Campaigns} \label{ssec: case study impersonation}

\textbf{Finding 21:} \textit{We discovered several campaigns impersonating well-known global brands so as to deceive end-users.} Such large-scale campaigns aim to mislead victims by using URLs containing text such as Apple and PayPal. For example, a large campaign (of size 4,081 unique URLs) used variations of domain and sub-domain names resembling the targeted Apple brand. The 4,081 unique URLs used a combination of 9 sub-domains, 12 domains, and 7 suffixes, with the ratio of the number of unique domains to the number of URLs standing at 0.29\%. The complete list of variations are:

\vspace{0.5em}

\noindent\fbox{%
    \parbox{\columnwidth}{%
        \textbf{Subdomains}: \texttt{www.apple.com}, \texttt{www.icloud.com}, \texttt{www.apple}, \texttt{icloud.com},  \texttt{apple}, \texttt{apple.com}, \texttt{icloud}, \texttt{www}, \texttt{mail}.\newline 
        \textbf{Domains}: \texttt{lcloud-com},  \texttt{online-support}, \texttt{findmy}, \texttt{get-apple}, \texttt{com-support}, \texttt{map-apple},  \texttt{id-info}, \texttt{wvvw-icloud}, \texttt{sign-in}, \texttt{viewlocation-icloud}, \texttt{map-log}, \texttt{com-fml}. \newline
        \textbf{Tld}: \texttt{us}, \texttt{in}, \texttt{support}, \texttt{live}, \texttt{review}, \texttt{com}, \texttt{mobi}.
    }%
}

\vspace{0.5em}

% \noindent\fbox{%
%     \parbox{\columnwidth}{%
%         \textbf{Subdomains}: \texttt{`www.apple.com'}, \texttt{`www.icloud.com'}, \texttt{`www.apple'}, \texttt{`icloud.com'},  \texttt{`apple'},  \texttt{`apple.com'}, \texttt{`icloud'}, \texttt{`www'}, \texttt{`mail'}. \newline 
%         \textbf{Domains}: \texttt{`lcloud-com'},  \texttt{`online-support'}, \texttt{`findmy'}, \texttt{`get-apple'}, \texttt{`com-support'}, \texttt{`map-apple'},  \texttt{`id-info'}, \texttt{`wvvw-icloud'}, \texttt{`sign-in'}, \texttt{`viewlocation-icloud'}, \texttt{`map-log'}, \texttt{`com-fml'}. \newline
%         \textbf{Tld}: \texttt{`us'}, \texttt{`in'}, \texttt{`support'}, \texttt{`live'}, \texttt{`review'}, \texttt{`com'}, \texttt{`mobi'}.
%     }%
% }

This shows a campaign that is precisely targeting the Apple brand. We found that the hash of URLs in this campaign were reported 104,311 times, and observed that an average 11 vendors were able to detect URLs from this campaign, which is significantly better performance compared to the results show in Figure~\ref{fig:positive distribution}. Surprisingly, we also observed that GSB only detected 2 of the URLs.

% This shows that the campaign is precisely targeting the Apple brand. We found that the hash of these URLs in this campaign was reported 104,311 times to VT, and observed that on average 11 security vendors are able to detect URLs from this campaign which is significantly better detection performance compared to the results show in Figure~\ref{fig:positive distribution}. Surprisingly, we also observed that GSB struggles in detecting these URLs, with only 2 out 4,081 URLs flagged by GSB (detection rate of only 0.04\%).

% This shows that the campaign is precisely targeting Apple brand. We found that the hash of these URLs in this campaign was reported 104,311 times to the VT.

% We observed that on average 11 security vendors are able to detect URLs from this campaign which is significantly better detection performance compared to the results show in Figure~\ref{fig:positive distribution}. Surprisingly, we also observed that GSB struggles in detecting these URLs -- only two out 4,081 URLs are flagged as malicious by GSB (detection rate in only 0.04\%) from the campaign.

\subsection{Malware Downloads} \label{ssec: case study malware}

We discovered a campaign attempting to download and install malware to victims' machines, with a size of 1,589 URLs and a source distribution of 16.57\%. The campaign used 261 unique domain names and a total of 1,589 URLs which were submitted 9,589 times. 1,175 of the URLs had the domain \texttt{ubar-}\texttt{pro4.ru}, which was confirmed as malicious~\cite{jeosandbox} and used for malware distribution.

% \subsection{Malware downloads} \label{ssec: case study malware}
% We discovered a phishing campaign which attempts to download and install malware to the victims' machines. The campaign size was 1,589 with source distribution of 16.57\%. The campaign used 261 unique domain names and total 1,589 URLs -- which were reported 9,589 times to the VT. Out of 1,589 URLs, 1,175 URLs had the domain  \texttt{http://ubar-pro4.ru}. We manually confirmed that this domain is malicious~\cite{jeosandbox} and is used to drop malware to victims' machines. 

\textbf{Finding 22:} \textit{We discovered close to 98\% (1,572/1,589) of URLs in the campaign communicates using transport layer protocol (TLS).} This finding is consistent with~\cite{sophos} which reports that nearly a quarter of malware now leverages TLS. Encryption is one of the strongest weapons used by attackers to obfuscate their code or URLs (specifically the path component of a URL) to remain undetected.

% \noindent \textbf{Finding 21:} \textit{We discovered nearly around 98\% (1,572/1,589) of URLs in the campaign communicates using transport layer protocol (TLS).} This finding is in consistence with~\cite{sophos} which reports that nearly a quarter of malware now communicates leverages TLS. Encryption is one of the strongest weapon leverages by attackers. They can use it to obfuscate their code or URLs (specifically Path parameter of a URL) to remain undetected by anti-malware services. 

On further analysis, we discovered that the attacker used the 1,572 URLs to download malware to victim machines via TLS. The downloaded files may have one of the following extension: \textit{.exe} (19 malicious executables found), \textit{.js} (28 javascript files), \textit{.zip} or \textit{.rar} (7 and 3 \textit{zip} and \textit{rar} files, respectively).

% After analysing the URLs in the campaign, we discovered that  attacker attempts to download malicious files to the victims' machines using TLS connections. We found that 1,572 URLs attempt to lure victims in downloading malicious files.
% The downloaded files may have one of the following extension: 1) \textit{.exe} (19 malicious executables found), \textit{.js} (28 javascript files), \textit{.zip} or \textit{.rar} (seven and three \textit{zip} and \textit{rar} files, respectively). 

We also discovered that URLs in this campaign are embedded with pointers to malicious torrent files. This is interesting as attackers no longer have to rely on downloading torrent files from a server but instead can do so directly from a torrent peer. This allows for strong persistence of malicious content as victim access is ensured even when trackers are closed, or down due to registration. We observed 46 (2.89\%) URLs with such capability, with very low detection rates for GSB and VT vendors. From the 1,589 unique URLs, only two were detected by GSB and on average, only one vendor flagged such URLs within this campaign.

% The main advantage for attackers is that the victim download the content of the torrent (malicious) from the URL, and even if the tracker is closed or down due to registration. 

% We also discovered that URLs in this campaign are embedded with pointers to torrent files (malicious). 
% This is interesting finding because attackers now do not have rely on downloading the torrent files from a web server, however, they can download it directly from a seeder/leecher. The main advantage of this for attacker is that the victim download the content of the torrent (malicious) from the URL, and even if the tracker is closed or down due to registration. We observed 46 (2.89\%) URLs with such capability. Finally, the detection rate for GSB and VT vendors were very low. From the 1,589 unique URLs, only two were detected by GSB and on average only one vendor flagged URLs in this campaign. 

% We observed that the detection rate for GSB and VT vendors is also very low -- out of unique 1,589 URLs in this campaign, only two were detected by GSB and on average only one vendor flagged URLs in this campaign. 

% \begin{table}[!ht]
% \centering
% \caption{URLs with unique characteristics.} \label{tab:url_charstcs}
% \resizebox{2.4in}{!}{
% \begin{tabular}{lr}
% \toprule
% \textbf{File type} & \textbf{Instances} \\ \midrule
% Encrypted files & 1,175 (73.94\%)\\
% \bottomrule
% \end{tabular}}
% \end{table}

\subsection{Fileless Malware} \label{ssec: case study fileless malware}

Some campaigns exhibit increased sophistication, with malware that leverages native legitimate and pre-installed system tools to execute an attack. Unlike traditional malware, fileless malware foregoes the need for malicious payloads which makes it harder to detect by disk scanning tools. Such malicious techniques that rely on using native tools is called ``living off the land''.

% Fileless malware is a type of sophisticated malware that leverages native, legitimate tools pre-installed into a system to execute an attack. Unlike traditional malware, fileless malware does not require an attacker to download malicious files on a victims’s machine -- making it hard to detect by disk scanning tools. Such fileless techniques that rely on using system's native tools to perform a malicious task is called ``living off the land''. 

\textbf{Finding 23:} \textit{We observed 23 campaigns that leverage system's pre-installed legitimate tools, such as PowerShell, to perform malicious tasks.}
The URLs within such campaigns are representative of a three-stage process: 1) The infection starts with phishing emails that are rigged with a URL to lure the victim to click on it. 2) Download a malicious payload using command-line tool (e.g., \textit{cmd.exe}) to invoke the PowerShell tool with code that is heavily obfuscated in order to evade analysis. The malicious file downloaded from remote server is kept in-memory rather than on-disk. 3) Finally, the file is executed which loads its plugins and then enters a communication loop which fetches tasks from command \& control servers.

% \noindent \textbf{Finding 22:} \textit{We observed that around 23 phishing campaigns leverage system's pre-installed legitimate tools, such as PowerShell, to perform malicious tasks.}
% We observed certain URLs are representative of a three-stage process: 1) The infection starts with phishing emails that are rigged with a URL to lure the victim to click on it. 2) Download a malicious payload using commandLine tool (e.g., \textit{cmd.exe}) to invoke PowerShell tool with code that is heavily obfuscated in order to evade analysis. The malicious file downloaded from remote server is kept memory rather than the disk. 3) Finally, the file is started to execute which loads its plugins and then enters a communication loop -- fetching commands from respective C2 server and dispatching them.

% Exploits are pieces of code or a sequences of commands that are employed by attackers using legitimate tools to perform malicious tasks.
% We observed that some campaigns used URLs that download and automatically ran malicious tasks using system's legitimate tools.

An example of such exploits is given below, where a clicked URL will download an executable (\textit{bad.exe}) from a remote server via PowerShell and cmd.exe, and then execute it:

% The following is an example URL (refined) that when clicked,  downloads malicious executable (\textit{bad.exe}) from remote server using PowerShell and cmd.exe tools.
% Once the file is downloaded, it is executed. 

\vspace{0.5em}

\noindent\fbox{%
    \parbox{\columnwidth}{%
    \textit{http://xxx.xxx.xxx.xxx/public/invokefunction\&function= call\_user\_func\_array\&vars[0]=system\&vars[1][]= \textbf{cmd.exe}$\backslash$/c$\backslash$\textbf{powershell}$\backslash$(new-object$\backslash$System.Net .WebClient). \textbf{DownloadFile}('www.bad.com/download.exe','$\backslash$SystemRoot$\backslash$ /Temp/\textbf{bad.exe}');\textbf{start}$\backslash$SystemRoot$ \backslash$/Temp/\textbf{bad.exe}}
    }%
}
\\[0.1em]

We also discovered campaigns that leverage other legitimate tools while performing malicious tasks. Examples of these include:

% We also discovered campaigns that leverage many additional legitimate tools while performing malicious tasks. For instance, the following is a list of legitimate tools leveraged by a campaign to carry-out malicious tasks.

\vspace{0.5em}

\noindent\fbox{%
    \parbox{\columnwidth}{%
    \textbf{Tools}: \textit{wmiprvse.exe, svchost.exe, conhost.exe, lsm.exe, sysstem.exe, lsass.exe, winlogon.exe, spoolsv.exe, csrss.exe, securityhealthservice.exe, services.exe, wininit.exe, cmd.exe, explorer.exe, wudfhost.exe, yourphone.exe, smss.exe, dwm.exe, taskhost.exe, csrss.exe, system.exe, regasm.exe, idle.exe}
    }%
}

%% file: Sections/7_Discussion.tex
\section{Discussion}

\textbf{Content Hash Detection}: Preceding sections have highlighted that traditional block-lists are easily improved if defenders consider the content hash of URLs. Previously unseen URLs could thus be rapidly identified in the absence of obvious malicious connections. We see evidence of this with cluster 16f3-21a6, where one URL (everyday-vouchers.com) remained unflagged for six submissions before being flagged by GSB on 2020-01-02. Shortly after, another URL (sweepstakehunter.com) bearing the same content hash was observed as remaining unflagged for 5 submissions (to end of data). 

% his would allow the detection of zero-day malicious URLs, where vendors cannot otherwise link suspicious URLs to existing ones.

% . On the seventh submission to VirusTotal on 2020-01-02, it was flagged as malicious by a single vendor (the same Google Safebrowsing).

A year on, \textit{both URLs} were marked as malicious by GSB from our secondary checks. Thus, a content-hash detector would have immediately flagged the second malicious URL as soon as it appeared within the VirusTotal platform in Jan 2020. This would have protected end-users from accessing it from an even earlier time.

It is noted that attackers could easily change generated hashes by automatically mutating malicious content for each end-user visit. This would evade content-hash detection as presumably, each URL results in a unique hash. In this case, vendors can fall back on methods such as linking IP addresses and server signatures. For phishing pages, HTML comparison methods such as in \cite{li2019stacking} could be used. Nonetheless, content hashes may still be of aid by initially clustering URLs, before more computationally intensive mechanisms are brought to bear. This could minimize the risk of false positives in more complex detectors deployed in later stages.

% However, this would still slow down attackers as they would need to implement evasive mechanisms.

% Another potential shortcoming is that attackers or vandals may use such techniques to perform indirect denial of service attacks. Consider if an attacker were to clone the Paypal login page completely, and deploy it to a server they control. That would have an identical hash signature to the legitimate Paypal page. They then set up an obvious phishing domain name and serve their cloned content. If vendors picked up the obvious phishing URL and thus the cloned page, automated process might add both to block-lists and the legitimate Paypal URL could be impacted.

\vspace{0.4em}

\noindent \textbf{Impact of Campaigns}: With over 20 findings presented, the community is better positioned to understand malicious URLs and thus better detect them. Traditional approaches must be improved as we cannot rely on the wholesale lumping of URLs into binary benign/malicious classifications. For deep-learning based detectors, active threat intelligence such as this could aid in teasing out learning features from disparate URLs that may have a common attacker. As attackers evolve their tactics, viewing the URLs as coherent campaigns opens up avenues for more refined defences. We also see the apparent limitations of current block-lists, which are not able to detect URLs that are obviously a part of malicious campaigns.

The insights and various case studies shared here exemplifies the tricks that attackers use to prolong and better target their campaigns. We see the range of domains, sub-domains and suffixes used to evade defences and deceive end-users in Sections 4.5 and 5.1. Malware payloads also exhibit a dangerous prevalence throughout the campaigns we discovered, with attackers becoming more apt at using standard security mechanisms to mask their payloads. We also see the efforts they go to in order to evade defences, with our findings on their use of widely variant URL lengths and propensity for longer URLs as seen in Section 4.4. It is hoped that such insights would be of use to the wider cyber-security community.

An issue of concern is that vendors are not able to effectively identify a swathe of malicious URLs. It would be unfair to expect high detection rates for all vendors represented on VirusTotal. However, there seems to be little agreement between vendors when URLs are indeed suspicious, evidenced by just 0.69\% of submissions being flagged by 11 or more vendors. The clear recommendation for the community is to not rely on the fact that there are a large number of vendors, as it may offer a false sense of completeness and security. Rather, care must be taken to focus only on a select few vendors.

Even so, researchers are advised that even the most performant vendor may struggle to detect whole campaigns. GSB, which is the premier block-list used by a large section of Internet users, suffers from poor detection rates in a variety of campaigns. This is illustrated in detail from the multiple findings within Sections 4 and 5. Our uncommon perspective from the lens of campaigns has laid bare the fact that, perhaps unsurprisingly, many malicious URLs are slipping thru block-lists and endangering end-users.

Finally, file-less malware is also a rising concern, with exploits encoded directly into the URLs themselves in order to compromise machines with and without secondary download steps. We postulate that the use of static code analysis tools may be of use here. That is, vendors may implement detectors that treat the URLs as lines of code in order to detect inline scripts that may harm users.

%% file: Sections/9_Conclusion.tex
\section{Conclusion}

While there has been extensive work studying malicious URLs, there remains a gap in the understanding of them as concerted \textit{campaigns}. The work herein is a culmination of extensive research and analysis into 311 million records submitted to VirusTotal from Dec 2019 to Jan 2020. Clusters of URLs are identified via their metadata and doubly verified as malicious before being analysed. This offers practical insights into the nature of campaigns along dimensions such as individual URL attributes, temporal aspects, and the victim brands they target. The nature of these campaigns are also explored with several case-studies that illustrate the techniques that victimise users and circumvent detection strategies.

% Evidence of \textit{connected campaigns} were also found, in which attackers are seen to change tactics as their operation progresses.

% URLs are central to a myriad of cyber-security threats, from phishing to the distribution of malware. Their inherent ease of use and wide familiarity is continuously abused by malicious actors to deceive end-users and evade even the most sophisticated defences.

Future work includes looking at measures to further group campaigns together. That is, to see if there are ways for us to connect different campaigns into a single large campaign that may be attributed to a common attacker. A recommendation we have for block-list providers is to simply block URLs that all point to the same content, if said content is proven to be malicious. This perhaps simple step can increase their protection coverage and leave them better able to detect zero-day URLs that appear as attacks evolve.

\section*{Acknowledgements}
The work has been supported by the Cyber Security Research Centre Limited whose activities are partially funded by the Australian Government’s Cooperative Research Centres Programme.

\noindent This work was supported by resources provided by the Pawsey Supercomputing Centre with funding from the Australian Government and the Government of Western Australia.

\noindent This research was supported by use of the Nectar Research Cloud and by the Tasmanian Partnership for Advanced Computing (TPAC).  The Nectar Research Cloud is a collaborative Australian research platform supported by the NCRIS-funded Australian Research Data Commons (ARDC).

% The work has been supported by the REDACTED whose activities are partially funded by the REDACTED.

% \noindent This work was supported by resources provided by the REDACTED with funding from the REDACTED and the REDACTED.

% \noindent This research was supported by use of the REDACTED and by the REDACTED. The REDACTED is a collaborative REDACTED research platform supported by the REDACTED-funded REDACTED.